\relax
\documentclass[letterpaper]{article} 
\usepackage{aaai21}  
\usepackage{times}  
\usepackage{helvet} 
\usepackage{courier}  
\usepackage[hyphens]{url}  
\usepackage{graphicx} 
\usepackage{natbib}  
\usepackage{caption} 

\usepackage{booktabs}
\usepackage{multirow}
\usepackage{float}
\usepackage{mathtools}
\usepackage{soul}
\usepackage{color}
\usepackage{enumitem}
\usepackage{hyperref}
\usepackage{diagbox}
\usepackage{makecell}
\usepackage[toc,page]{appendix}
\usepackage[colorinlistoftodos]{todonotes}
\usepackage{subfigure}
\usepackage{algpseudocode}
\usepackage{amsmath}
\usepackage{amsfonts}

\usepackage[misc]{ifsym}

\usepackage{subfigure}
\usepackage{graphicx}
\usepackage{booktabs}
\usepackage{multirow}
\usepackage{float}
\usepackage{mathtools}
\usepackage{soul}
\usepackage{color}
\usepackage{enumitem}
\usepackage{hyperref}
\usepackage{diagbox}
\usepackage{makecell}
\usepackage[toc,page]{appendix}
\usepackage[colorinlistoftodos]{todonotes}
\usepackage{subfigure}
\usepackage{algpseudocode}
\usepackage{amsmath}
\usepackage{amssymb}
\usepackage{mathtools}
\usepackage{bm}
\usepackage{setspace}
\usepackage[ruled,linesnumbered]{algorithm2e}

\usepackage[switch]{lineno}

\usepackage{appendix}

\newtheorem{myDef}{Definition}

\hypersetup{hypertex=true,
            colorlinks=true,
            linkcolor=black,
            anchorcolor=black,
            citecolor=black}

\newcommand{\PreserveBackslash}[1]{\let\temp=\\#1\let\\=\temp}
\newcolumntype{C}[1]{>{\PreserveBackslash\centering}p{#1}}
\newcolumntype{R}[1]{>{\PreserveBackslash\raggedleft}p{#1}}
\newcolumntype{L}[1]{>{\PreserveBackslash\raggedright}p{#1}}

\usepackage[colorinlistoftodos]{todonotes}
\newcommand{\nop}[1]{}

\title{Connecting Latent ReLationships over Heterogeneous Attributed Network \\
for Recommendation}
\author{
		Ziheng Duan\textsuperscript{\rm 1, \rm 2}       		
		Yueyang Wang\textsuperscript{\rm 2}\thanks{Corresponding Author.}
\\
		Weihao Ye\textsuperscript{\rm 1}
		Zixuan Feng\textsuperscript{\rm 2}
		Qilin Fan\textsuperscript{\rm 1}
		Xiuhua Li\textsuperscript{\rm 1}
\\
}
\affiliations{
    \textsuperscript{\rm 1}School of Big Data and Software Engineering, Chongqing University, Chongqing 401331, China\\
    \textsuperscript{\rm 2}Department of Control Science and Technology, Zhejiang University, Hangzhou 310027, China\\

    duanziheng@zju.edu.cn, yueyangw@cqu.edu.cn\\
    yeweihao@cqu.edu.cnn, fengzixuan121@gmail.com, fanqilin@cqu.edu.cn, lixiuhua1099@gmail.com\\
}
\begin{document}

\maketitle

\begin{abstract}
\begin{quote}
Recently, deep neural network models for graph-structured data have been demonstrating to be influential in recommendation systems.
Graph Neural Network (GNN), which can generate high-quality embeddings by capturing graph-structured information, is convenient for the recommendation.
However, most existing GNN models mainly focus on the homogeneous graph.
They cannot characterize heterogeneous and complex data in the recommendation system.
Meanwhile, it is challenging to develop effective methods to mine the heterogeneity and latent correlations in the graph.
In this paper, we adopt Heterogeneous Attributed Network (HAN), which involves different node types as well as rich node attributes, to model data in the recommendation system.
Furthermore, we propose a novel graph neural network-based model to deal with HAN for Recommendation, called HANRec.
In particular, we design a component connecting potential neighbors to explore the influence between neighbors and provide two different strategies with the attention mechanism to aggregate neighbors’ information.
The experimental results on two real-world datasets prove that HANRec outperforms other state-of-the-arts methods\footnote{Code will be available when this paper is published.}.
\end{quote}
\end{abstract}

\section{Introduction}
\begin{figure*}
  \includegraphics[width=1\textwidth]{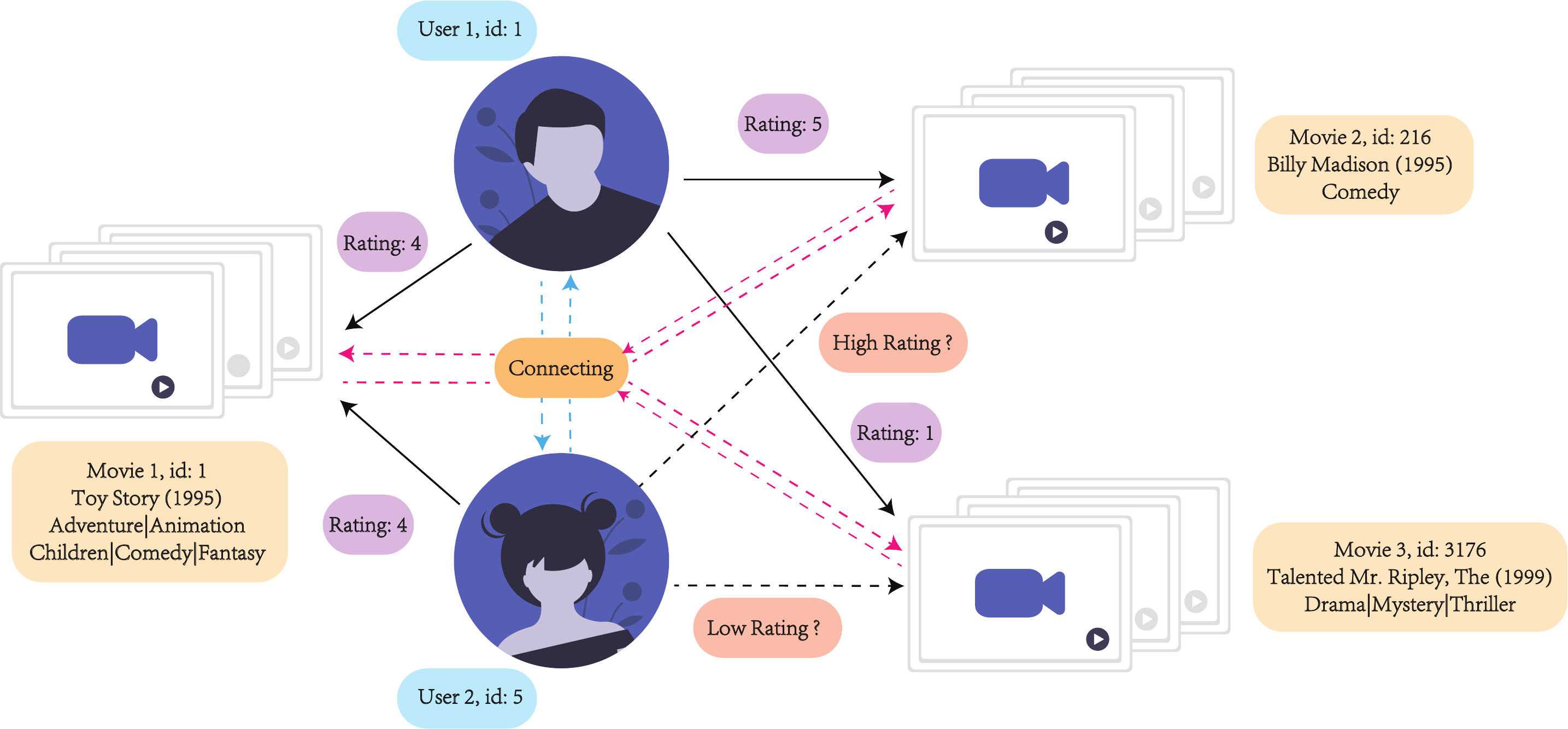}
\caption{An example of connecting the users and movies from the real-world dataset: MovieLen.
The solid black line represents the rating scores of movies by users.
We use the blue dashed line to indicate the potential relationships between users and the red dashed line to demonstrate the potential relationships between movies through the connect module.
The black dotted line is the rating score of the movie 2 and 3 of user 2, which we want to predict.}
\label{fig:intro}
\end{figure*}

As it becomes more convenient for users to generate data than before, mass data appeared on the Internet are no longer with simple structure but usually with more complex types and structures.
This makes the recommendation system \cite{ricci2011introduction}, which helps users discover items of interest, attract widespread attention, and face more challenges \cite{lekakos2008hybrid}.
Traditional recommendation methods, such as matrix factorization \cite{ma2008sorec}, are designed to explore the user-item interactions and generate an efficient recommendation.
However, because of calculating each pair of interactions, they have difficulty processing sparse and high-volume data.
Furthermore, it is more challenging to deal with heterogeneous and complex data derived from different sources \cite{wu2015trust, kefalas2018recommendations}.

As the data describing users and items in the recommendation system can be naturally organized as a graph structure.
The graph-based deep learning methods \cite{kipf2016semi} have been applied in recommendation fields \cite{gnnforsocialrecommendation} and alleviate the enormous amount and the sparsity problem to some extent.
Meanwhile, heterogeneous attributed network (HAN) \cite{wang2019heterogeneous}, consisting of different types of nodes representing objects, edges denoting the relationships, and various attributes, is a unique form of graph and is powerful to model heterogeneous and complex data in the recommendation system \cite{zhu2021recommending}.
Therefore, designing a graph-based recommendation method to mine the information in HAN is a promising direction \cite{zhong2020hybrid}.

Recently, Graph Neural Network (GNN) \cite{defferrard2016convolutional}, a kind of deep learning-based method for processing graphs, has become a widely used graph analysis method due to its high performance and interpretability.
GNN based recommendation methods have proved to be successful to some extent because they could simultaneously encode the graph structure and the attributes of nodes \cite{derr2018signed}.
However, the advances of GNNs are primarily concentrated on homogenous graphs, so they still encounter limitations to utilize rich information in HAN.
The major challenge is caused by the heterogeneity of the recommendation graph.
The recommendation methods should be able to measure similarities between users and items from various aspects \cite{gnnforsocialrecommendation}.
On the one hand, different types of nodes have various attributes that cannot be directly applied due to the different dimensions of node attributes \cite{wang2019heterogeneous}.
For example, in movie rating graphs, a "user" is associated with attributes like interests and the number of watched movies, while a "movie" has attributes like genres and years.
On the other hand, the influence of distinct relationships is different, which should not be treated equally \cite{chen2020learning}.
For instance, the correlation between "user-Rating-movie" is intuitively stronger than "movie-SameGenres-movie."
Hence, how to deal with different feature spaces of various objects' attributes and how to make full use of heterogeneity to distinguish the impact of different types of entities are challenging.

Furthermore, GNNs are based on the connected neighbor aggregation strategy to update the entity's representation \cite{zhou2018graph}.
However, the reality is that some potential relationships of entities are not directly connected, but implicit \cite{wu2020connecting}.
For example, as shown in Fig. \ref{fig:intro}, both user 1 and user 2 score the movie 1 as the same rating 4.
That in part reflects the interests of the two users are similar. Analogously, movie 1 and movie 2 have the same genres.
This infers there was some potential relationship between the two movies. Nevertheless, these implicit relationships are hardly captured by GNNs.
Meanwhile, the degree of rating could reflect the preferences of users.
For instance, user 1 rates the movie 2 as 5, but the movie 3 as 1.
This may infer that user 2 prefers movie 3 to movie 2.
Therefore, how to explicitly model entities with potential relationships to provide references for each other's recommendation and distinguish the rating weights are significant.

To better overcome these challenges mentioned above, we propose a neural network model to deal with Heterogeneous Attributed Network for Recommendation, called HANRec, which can make full use of the heterogeneity of graphs and deeper encode the latent relationships.
Specifically, we first design a component connecting potential neighbors to explore the influence between neighbors with potential connections.
The connecting component also assigns users' rating information to different weights and integrates them into the potential relationships.
Next, we design homogeneous aggregation and heterogeneous aggregation strategies to aggregate feature information of both connected neighbors and potential neighbors.
It is worth mentioning that we introduce the attention mechanism \cite{vaswani2017attention} to measure the different impacts of heterogeneous nodes. The learned parameters in the two aggregations are different so that they can model different patterns of the heterogeneous graph structure.
Finally, we use the entities' high-quality embeddings to make corresponding recommendations through the recommendation generation component.
To summarize, we make the following contributions:

\begin{itemize}
\item We propose a new method to connect the users, thus providing a potential reference for recommendations, where existing methods usually miss.
\item We present a novel framework for the recommendation, making full use of the heterogeneity in graphs and design strategies for gathering information for entities of different types.
\item We design an attention mechanism to characterize the different influences among entities.
\item Our model outperforms previous state-of-the-art methods in two recommendation tasks.
\end{itemize}

\section{Related Work}
\label{sec: Related work}
This section briefly introduces some related works on recommendation algorithms, mainly from traditional methods and deep learning methods.

For the past years, the use of social connections for the recommendation has attracted great attention \cite{10.1145/2872427.2882971,7558226}.
Traditional methods to deal with recommendation problems mainly include content-based recommendation algorithms and collaborative filtering algorithms.
The basic idea of Content-Based Recommendation (CB) \cite{10.5555/1768197.1768209} is to extract product features from known user preference records and recommend the product that is most similar to his/her known preference to the user.
The recommendation process of CB can be divided into three steps: Item Representation, Profiles Learning, and Generate Recommendation Generation.
A further complication in generating a product representation is that the information for the product is unstructured. 
The most typical unstructured product information is text. 
The weight of words can be calculated by using Term Frequency (TF) \cite{Ramos2003UsingTT} and Inverse Document Frequency (IDF) \cite{Ramos2003UsingTT} to vectorize text.
Feature learning is the process of modeling user interests.
The training process of all user interest models is a supervised classification problem that can be solved using traditional machine learning algorithms. 
For example, decision tree algorithm \cite{97458}, k-nearest neighbor algorithm \cite{6313426}, Naive Bayes algorithm \cite{article}, support vector machine \cite{885819}, etc. 
CB uses different ways to generate recommendations depending on the recommendation task scenario. 
This strategy works well in situations where the product's attributes can be easily identified.
Nevertheless, it is easy to over-specialize the user's interest, and it is challenging to find out the user's potential interest.
The idea of Collaborative Filtering (CF) \cite{10.1145/963770.963772} is to find some similarity through the behaviors of groups and make recommendations for users based on this similarity. 
The recommendation process is: establishing a user behavior matrix, calculating similarity, and generating recommendations.
Collaborative filtering based on matrix decomposition decomposes the user rating matrix into user matrix U and product matrix V. 
It then recombines U and V to get a new user rating matrix to predict unknown ratings. 
Matrix decomposition includes Funk-SVD model \cite{koren2009matrix}, PMF model \cite{mnih2007probabilistic}, etc.

Deep learning, as an emerging method, also has many achievements of learning on graph structure data \cite{7974879}.
The purpose of graph representation learning is to find a mapping function that can transform each node in the graph into a low-dimensional potential factor.
The low-dimensional latent factors are more efficient in calculations and can also filter some noise.
Based on the nodes' latent factors in the graph, machine learning algorithms can complete downstream tasks more efficiently, such as recommendation tasks and link prediction tasks.
Graphs can easily represent data on the Internet, making graph representation learning more and more popular.
Graph representation learning includes random walk-based methods and graph neural network-based methods.
The random walk-based methods sample the paths in the graph, and the structure information near the nodes can be obtained through these random paths.
For example, the DeepWalk proposed by Perozzi et al. \cite{10.1145/2623330.2623732} applies the ideas of Natural Language Processing (NLP) to network embedding.
DeepWalk treats the relationship between the user and the product as a graph, generating a series of random walks.
A continuous space of lower latitudes represents the user vector and the product vector.
In this graph representation space, traditional machine learning methods, such as Logistic Regression (LR), can predict users' ratings of products to obtain more accurate results.
Collaborative Deep Learning (CDL) proposed by Wang et al. \cite{10.1145/2783258.2783273} jointly deals with the representation of product content information and users' rating matrix for products.
CDL relies on user reviews of the product and information about the product itself.

Some researchers have also used graph neural networks (GNNs) to complete recommendation tasks in recent years.
Graph Convolutional Network \cite{kipf2016semi} (GCN) is one of the representative methods.
GCN uses convolution operators on the graph to iteratively aggregate the neighbor embeddings of nodes.
It utilizes the Laplacian matrix of the graph to implement the convolution operation on the topological graph.
In the multi-layer GCNs, each convolutional layer processes the graph's first-order neighborhood information. 
Superimposing multiple convolutional layers can realize information transfer on the multi-level neighborhood.
On this basis, some GNN-based frameworks for recommendation tasks have been proposed.
According to whether to consider the order of items, recommendation systems can be divided into regular recommendation tasks, and sequential recommendation tasks \cite{adomavicius2005toward,kang2018self}.
The general recommendation considers users to have static interest preferences and models the degree of matching between users and items based on implicit or explicit feedback.
GNN can capture user-item interactions and learn user and item representations.
The sequential recommendation captures the serialization mode in the item sequence and recommends the next item of interest to the user. There are mainly methods based on Markov chain (MC) \cite{rendle2010factorizing, he2016fusing}, based on RNN \cite{hidasi2018recurrent,tan2016improved}, and based on attention and self-attention mechanisms \cite{liu2018stamp,li2017neural}.
With the advent of GNN, some works convert the item sequence into a graph structure and use GNN to capture the transfer mode.
Since this paper focuses on discussing the former, we mainly introduce the work of the general recommendation.

General recommendation uses user-item interaction to model user preferences.
For example, GC-MC \cite{berg2017graph} deals with rating score prediction, and the interactive data is represented as a bipartite graph with labeled edges. 
GC-MC only uses interactive items to model user nodes and ignores the user's representation.
So the limitations of GC-MC are:
1) it uses mean-pooling to aggregate neighbor nodes, assuming that different neighbors are equally important;
2) it only considers first-order neighbors and cannot make full use of the graph structure to spread information.
Online social networks also develop rapidly in recent years.
Recommendation algorithms that use neighbors' preferences to portray users' profiles have been proposed, which can better solve the problem of data sparsity and generate high-quality embeddings for users.
These methods use different strategies for influence modeling or preference integration.
For instance, DiffNet \cite{wu2019neural} models user preferences based on users' social relationships and historical behaviors, and it uses the GraphSAGE framework to model the social diffusion process. 
DiffNet uses mean-pooling to aggregate friend representations and mean-pooling to aggregate historical item representations to obtain the user's representation in the item space.
DiffNet can use GNN to capture a more in-depth social diffusion process. However, this model's limitations include: 1) The assumption of the same influence is not suitable for the real scene; 2) The model ignores the representation of the items and can be enhanced by interactive users.
GraphRec proposed by Fan et al. \cite{10.1145/3308558.3313488} learns the low-dimensional representation of users and products through graph neural networks. 
GraphRec used the mean square error of predicted ratings to guide the neural network's parameter optimization and introduced users' social relationships and attention mechanisms to describe user preferences better.
With the emergence of heterogeneous and complex data in the recommendation network, some work has introduced knowledge graphs (KG) into recommendation algorithms.
The challenge of applying the KG to recommendation algorithms comes from the complex graph structure, multiple types of entities, and relationships in KG. 
Previous work used KG embedding to learn the representation of entities and relationships; or designed meta-path to aggregate neighbor information. Recent work uses GNN to capture item-item relationships.
For example, KGCN \cite{wang2019knowledge} uses user-specific relation-aware GNN to aggregate neighbors' entity information and uses knowledge graphs to obtain semantically aware item representations.
Different users may impose different importance on different relationships. 
Therefore, the model weights neighbors according to the relationship and the user, which can characterize the semantic information in the KG and the user's interest in a specific relationship.
The item's overall representation is distinguished for different users, and semantic information in KG is introduced.
Finally, predictions are made based on user preferences and item representation.
IntentGC \cite{zhao2019intentgc} reconstructs the user-to-user relationship and the item-to-item relationship based on the multi-entity knowledge graph, greatly simplifying the graph structure.
The multi-relationship graph is transformed into two homogeneous graphs, and the user and item embeddings are learned from the two graphs, respectively.
This work also proposes a more efficient vector-wise convolution operation instead of the splicing operation.
IntentGC uses local filters to avoid learning meaningless feature interactions (such as the age of one's node and neighbor nodes' rating).

Although the previous work has achieved remarkable success, the exploration of potential relationships in social recommendations has not been paid enough attention. 
In this paper, we propose a graph neural network framework that can connect potential relationships in the network to fill this gap.

\section{Proposed Framework} 
\label{sec: Proposed Framework}
\begin{figure*}
\centering
  \includegraphics[width=1\textwidth]{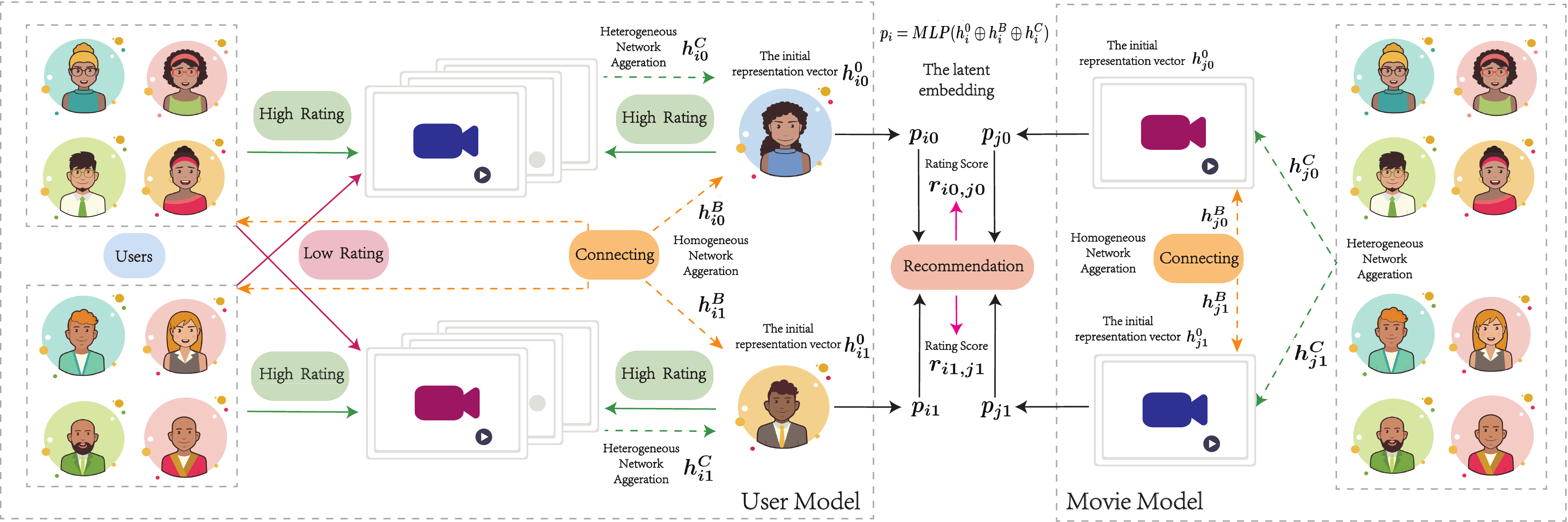}
\caption{The architecture of HANRec. It contains four major components: connecting potential neighbors, homogeneous aggregation, heterogeneous aggregation and rating generation.}
\label{fig:HANRec}
\end{figure*}

In this section, we present the proposed HANRec in detail.
First, we define the heterogeneous attributed network, the formula expression of the recommendation and link prediction problem, and the symbols used in this paper.
Later, we give an overview of the proposed framework and the details of each component.
Finally, we introduce the optimization objectives of HANRec.
The schematic of connecting the users and the model structure is shown in Fig. \ref{fig:HANRec}.

\begin{table*}[t]
\renewcommand\arraystretch{1.5}
\centering
\caption{Symbols' Definition and Description}
\scalebox{1}{
\begin{tabular}{c|c}
\hline
Symbols                 & Definitions                                                                                                                                         \\ \hline

$B(v_i)$                 & The set of entities of the same type as entity $v_i$.                                                     \\ \hline

$C(v_i)$                 & The set of entities of different types as entity $v_i$.                                                    \\ \hline

$N(v_i)$                 & The set of neighbor entities of entity $v_i.$                                                                                                               \\ \hline

$N'(v_i)$                & The set of entities that have potential connections with entity $v_i$.                              \\ \hline

$r_{i,j}$                   & The relationship between entity $v_i$ and $v_j$.                                                                                                         \\ \hline

$p_i$                 
& The embedding vector of entity $v_i$.                                                                                                                  
 \\ \hline

$e_{i,j}$                   & The rating embedding for the rating level (entity $v_i$ to $v_j$).                                \\ \hline

$d$                    & The embedding dimension.                                                                                                                                                                                   \\ \hline

$h_i^0$ & The initial representation vector of entity $v_i.$ \\ \hline

$h_i^B$ & The influence embedding among entity $v_i$ and entities of the same type. \\ \hline

$h_i^C$ & The influence embedding among entity $v_i$ and entities of different types. \\ \hline

$f_{i,j}$                   & The opinion-aware interaction representation between entity $v_i$ and $v_j$.                         \\ \hline

$\alpha^*_{i,j}$                    & Attention parameter between entity of the same type ($v_j$ to $v_i$).                                      
 \\ \hline

$\beta^*_{i,j}$                    & Attention parameter between entity of different types ($v_j$ to $v_i$).
 \\ \hline

$\alpha_{i,j}$                    & The normalized attention parameter between entity of the same type ($v_j$ to $v_i$).
 \\ \hline

$\beta_{i,j}$                    & The normalized attention parameter between entity of different types ($v_j$ to $v_i$).
 \\ \hline

$\bm{W}, \bm{b}$                    & The weight and bias in neural networks.
 \\ \hline
\end{tabular}
}
\label{notation}
\end{table*}

\subsection{Problem Formulation and Symbols Definition}

\begin{myDef}
    Heterogeneous Attributed Network
\end{myDef}

A heterogeneous attributed network can be represented as $G = (V, E, A)$.
$V$ is a set of nodes representing different types of objects, $E$ is a set of edges representing relationships between two objects, and $A$ denotes the attributes of objects.
The heterogeneity of the graph is reflected as: $type(nodes) + type(edges) > 2$.
On the other hand, when $type(nodes) = type(edges) = 1$, we call it a homogeneous graph \cite{cen2019representation}.
Attributed graphs mean that each node in graphs has corresponding attributes.
Entities with attributes are widespread in the real world.
For example, in a movie recommendation network, each movie has its genres and users have their preferences; in an author-paper network, each author and paper has related research topics.
Therefore, most networks in the real world exist as heterogeneous attribute graphs, which is the focus of this paper.

\begin{myDef}
    Recommendation
\end{myDef}

In this paper, the task of recommendation is focused.
For a graph $G = (V, E)$, $v_i$ and $v_j$ are two entities in this graph.
In the triple ($r$, $v_i$, $v_j$), $r$ represents the relationship between $v_i$ and $v_j$.
A recommendation model learns a score function, and outputs the relationship $r$ between $v_i$ and $v_j$:
$r_{i,j} = Recommendation\_Model(v_i, v_j)$.
For example, in the user-movie-rating recommendation network, the main focus is to recommend movies that users might be interested in.
This requires the recommendation model to accurately give users the possible ratings of these movies (commonly a 5-point rating, etc.).
So in most cases, this is a regression problem.

\begin{myDef}
    Link Prediction
\end{myDef}
We also discussed link prediction in this paper.
For a graph $G = (V, E)$, $v_i$ and $v_j$ are two entities in this graph.
In the triple ($r$, $v_i$, $v_j$), $r$ represents the relationship between $v_i$ and $v_j$.
A link prediction model learns a score function, and outputs the relationship $r$ between $v_i$ and $v_j$:
$r_{i,j} = Link\_Prediction\_Model(v_i, v_j)$.
In most cases, the value of $r$ is 0 or 1.
This means that there is no edge or there is an edge between the two entities, which is a binary classification problem.

The purpose of link prediction is to infer missing links or predict future relations based on the currently observed part of the network. 
This is a fundamental problem with a large number of practical applications in network science.
The recommendation problem can also be regarded as a kind of link prediction problem. 
The difference is that recommendation is a regression task that needs to predict specific relationships (such as the ratings of movies by people in a movie recommendation network). 
At the same time, link prediction is a binary classification task, where we need to determine whether there is a link connected between the two entities.

The mathematical notations used in this paper are summarized in Table \ref{notation}.

\subsection{An Overview of HANRec}
HANRec consists of four components: Connecting Potential Neighbors, Homogeneous Network Aggregation, Heterogeneous Network Aggregation, and Recommendation Generation.
Connecting Potential Neighbors aims to fully explore the influence between neighbors with potential connections.
Like the common user-movie-rating network, there is a lot of data for users' ratings of movies, but there is little or no interaction between users.
However, the interaction between users is important, and there may be potential influences between users.
Users with the same interests will have a high probability of giving similar ratings to the same movie.
This component uses shared entities to generate a connection path, which utilizes scores and features to model the potential influence and generate high-quality entity representations.

Homogeneous Network Aggregation is responsible for aggregating information of entities of the same type and quantifying the influence of different neighbors on the entity through an attention mechanism.
This component is mainly to portray the influence between entities of the same type.
Heterogeneous Network Aggregation is responsible for aggregating information of entities of different types.
We also use the user-movie-rating network to illustrate.
Different movie genres and the user's ratings will describe the user's preferences from a particular perspective.
Heterogeneous Aggregation Network will focus on describing an entity's characteristics from the perspective of entities of different types.
Recommendation generation gathers all the information related to the entity and generates its high-quality embedding.
This component also matches the embeddings of other entities to make the final recommendations.

As shown in Fig. \ref{fig:HANRec}, in the movie recommendation network, we connect different users/movies through the designed connection method (as indicated by the orange arrow).
Then we use homogeneous and heterogeneous aggregation to combine users/movies its own characteristics and generate the final representation embedding.
Finally, the user's embedding is compared with the movie's embedding to be evaluated, and the final rating score is generated.
Next, we introduce the details of each component.

\subsection{Connecting Potential Neighbors}
We design this component to fully explore the influence between neighbors with potential connections. 
Note that in the common user-movie-rating network, there may not be a direct link among users and users, movies and movies.
At this time, we can connect the users and movies through second-order neighbors.
Suppose that $v_j \in B(v_i)\cap \overline{N(v_i)}$, $v_k \in C(v_i)\cap N(v_i)$, $v_k \in C(v_j)\cap N(v_j)$,
where $B(v_i)$ is the set of entities that are the same type of $v_i$; 
$C(v_i)$ is the set of entities that are different types of $v_i$ and $N(v_i)$ represents the set of neighbors of $v_i$.
We can infer the influence of $v_j$ on $v_i$ from the common neighbor $v_k$:
\begin{equation}
\label{eq:connect}
f_{i,j} = MLP(h_k^0 \oplus e_{i,k} \oplus h_j^0 \oplus e_{k,j}),
\end{equation}
where $f_{i.j}$ is the opinion-aware interaction representation between entity $v_i$ and $v_j$; 
$MLP$ means Multilayer Perceptron \cite{pal1992multilayer};
$\oplus$ represents the concatenate operator of two vectors;
$h_k^0$ and $h_j^0$ represent the initial features of $v_k$ and $v_j$, respectively;
$e_{i,k}$ and $e_{k,j}$ represent the rating embedding for the rating level between $v_i$, $v_k$ and $v_k$, $v_j$, respectively.
Through this connecting way, we use $f_{i,j}$ to represent their previous potential relationship between $v_i$ and $v_j$.
Then $v_j$ is added to $N'(v_i)$, which represents the set of entities that have potential connections with entity $v_i$.
The connection method we designed can fully integrate the characteristics of entities and the relationships between entities in the path.
For example, there may not be a direct connection between users in a common movie recommendation network. 
In this way, the two users can be connected by using the movies they have watched together. 
The movie genres and the user's rating of the movie can provide an essential reference for users' potential relationships.

\subsection{Homogeneous Network Aggregation}
Homogeneous Network Aggregation aims to learn the influence embedding $h_i^B$, representing the relationship among entity $v_i$ and entities of the same type.
Given the initial representation of entity $v_j$, and the rating embedding $e_{i,j}$, the opinion-ware interaction representation $f_{i,j}$ can be expressed as:
\begin{equation}
f_{i,j} = MLP(h_j^0 \oplus e_{i,j}).
\end{equation}
The attention parameter $\alpha^*_{i,j}$ reflecting the influence of different entities $v_i$ and $v_j$ is designed as:
\begin{equation}
\alpha^*_{i,j} = \bm{w^T_2}\cdot \sigma(\bm{W_1} \cdot [h_i^0 \oplus h_j^0] + \bm{b_1}) + \bm{b_2},
\end{equation}
where $\bm{W}$ and $\bm{b}$ represents the weight and bias in neural networks.
The normalized attention parameter $\alpha_{i,j}$ is as follows:
\begin{equation}
\alpha_{i,j} = \frac{exp(\alpha^*_{i,j})}{\sum_{k\in B(v_i)\cap (N(v_i)\cup N'(v_i))}exp(\alpha^*_{i,k})}.
\end{equation}
The design of attention parameters is based on the assumption that entities with similar characteristics should have greater influence among them.
And we use a neural network to learn this influence adaptively.
In this way, we get the information gathered by entity $v_i$ from its homogeneous graph:
\begin{equation}
h_i^B = \sum_{j\in B(v_i)\cap (N(v_i)\cup N'(v_i))}(\alpha_{i,j} * f_{i,j})
\end{equation}
It is worth mentioning that when we aggregate neighbor information, we not only consider neighbors with edge connections but also consider neighbors with potential relationships discovered through our connection method.
We aggregate their influence on entity $v_i$ in the meantime.
Subsequent experiments proved the superiority of our design ideas.

\subsection{Heterogeneous Network Aggregation}
Heterogeneous Network Aggregation aims to learn the influence embedding $h_i^C$, representing the relationship among entity $v_i$ and entities of different types.
Given the initial representation of entity $v_j$, and the rating embedding $r_{i,j}$, the opinion-ware interaction representation $f_{i,j}$ can be expressed as:
\begin{equation}
f_{i,j} = MLP(h_j^0 \oplus e_{i,j})
\end{equation}
The attention parameter $\beta^*_{i,j}$ reflecting the influence of different entities $v_i$ and $v_j$ is designed as:
\begin{equation}
\beta^*_{i,j} = \bm{w^T_4}\cdot \sigma(\bm{W_3} \cdot [h_i^0 \oplus h_j^0] + \bm{b_3}) + \bm{b_4}
\end{equation}
The normalized attention parameter $\beta_{i,j}$ is as follows:
\begin{equation}
\beta_{i,j} = \frac{exp(\beta^*_{i,j})}{\sum_{k\in C(v_i)\cap (N(v_i)\cup N'(v_i))}exp(\beta^*_{i,k})}
\end{equation}
In this way, we get the information gathered by entity $v_i$ from its heterogeneous graph:
\begin{equation}
h_i^C = \sum_{j\in C(v_i)\cap (N(v_i)\cup N'(v_i))}(\beta_{i,j} * f_{i,j})
\end{equation}
According to the different types of neighbors, we designed two aggregation strategies to highlight the impact of different types of entities on $v_i$.
For example, in a common movie recommendation network, the influence of movies and other people on a person should be different, and mixing them cannot make full use of the graph's heterogeneity.
The subsequent experimental part also proved our conjecture.

\subsection{Recommendation Generation}
After gathering information from entities of the same and different types, we can easily get the latent representation of entity $v_i$:
\begin{equation}
p_i = MLP(h_i^0 \oplus h_i^B \oplus h_i^C),
\end{equation}
where $h_i^0$ is the initial representation of entity $v_i$, such as the user's preferences in the user-movie-rating network, the movie genres, etc.
$h_i^B$ and $h_i^C$ represent the influence embeddings indicating the relationships among entity $v_i$ and entities of the same type and different types, respectively.
For $v_i$ and $v_j$, after getting the embedding that aggregates a variety of information ($p_i$ and $p_j$), their relationship can be measured as:
\begin{equation}
r_{i,j} = MLP(p_i \oplus p_j).
\end{equation}
So far, the entire end-to-end recommendation prediction process has been completed.

\IncMargin{1em}
\begin{algorithm*} 
\begin{spacing}{0.8}
\SetKwData{Left}{left}
\SetKwData{This}{this}
\SetKwData{Up}{up} 
\SetKwFunction{Union}{Union}
\SetKwFunction{FindCompress}{FindCompress} \SetKwInOut{Input}{Input}
\SetKwInOut{Output}{Output}

\Input{The heterogeneous attributed network $G = (V, E, A)$; two entities $v_1$ and $v_2$ whose relationship needs to be evaluated} 
\Output{The relationship $r_{1,2}$ between the two entities $v_1$ and $v_2$}
\BlankLine 

\emph{$P = \{\}$}\; 
\tcp{$P$ is used to record the representation embedding of each entity $v$}
\For{$v_i$ in \{$v_1$, $v_2$\}}{ 
    \emph{$N'(v_i) = \{\}$}\;
    \tcp{$N'(v_i)$ records the entities that have potential connections with entity $v_i$}
    \For{$v_m$ in $N(v_i)$}{
        $f_{i,m} = MLP(h_m^0\oplus e_{i,m})$\;
        \tcp{The opinion-aware interaction representation between entity $v_i$ and $v_m$}
        \For{$v_n$ in $N(v_m)$}{
            \If{$v_n \neq v_i$}
            {
                $N'(v_i) \leftarrow v_n$\;
                \tcp{Add $v_n$ to $N'(v_i)$}
                $f_{i,n} = MLP(h_m^0\oplus e_{i,m}\oplus h_n^0 \oplus e_{m,n})$\;
                \tcp{The opinion-aware interaction representation between entity $v_i$ and $v_n$}
            }
        }
    }
    \tcp{Connecting Potential Neighbors}
			$\alpha^*_{i,j} = \bm{w^T_2}\cdot \sigma(\bm{W_1} \cdot [h_i^0 \oplus h_j^0] + \bm{b_1}) + \bm{b_2}$,
			$\alpha_{i,j} = \frac{exp(\alpha^*_{i,j})}{\sum_{k\in B(v_i)\cap (N(v_i)\cup N'(v_i))}exp(\alpha^*_{i,k})}$\;
    \tcp{The attention parameters in homogeneous aggregation}
    $h_i^B = \sum_{v_k\in B(v_i)\cap (N(v_i)\cup N'(v_i))}(\alpha_{i,k} * f_{i,k})$\;
    \tcp{Homogeneous Aggregation}
    $\beta^*_{i,j} = \bm{w^T_4}\cdot \sigma(\bm{W_3} \cdot [h_i^0 \oplus h_j^0] + \bm{b_3}) + \bm{b_4}$,
			$\beta_{i,j} = \frac{exp(\beta^*_{i,j})}{\sum_{k\in C(v_i)\cap (N(v_i)\cup N'(v_i))}exp(\beta^*_{i,k})}$\;
    \tcp{The attention parameters in heterogeneous aggregation}
    $h_i^C = \sum_{v_k\in C(v_i)\cap (N(v_i)\cup N'(v_i))}(\beta_{i,k} * f_{i,k})$\;
    \tcp{Heterogeneous Aggregation}
    
  $p_i$ = $MLP(h_i^0 \oplus h_i^B \oplus h_i^C)$\;
  \tcp{Fuse $v_i$'s initial feature and the information aggregated from homogeneous and heterogeneous neighbors to obtain the final representation embedding $p_i$}
  $P \leftarrow p_i$\;
  \tcp{Add p to P}
} 
$r_{1,2}$ = $MLP(p_1 \oplus p_2)$\;
\tcp{Compute the relationship $r_{1,2}$ between two entity $v_1$ and $v_2$}
\Return{$r_{1,2}$}
\end{spacing}
\caption{HANRec algorithm framework}
\label{algorithm}
\end{algorithm*}
\DecMargin{1em} 

\subsection{Objective Function}
To optimize the parameters involved in the model, we need to specify an objective function to optimize.
Since the task we focus on in this work is rating prediction and link prediction, the following loss function is used referred to \cite{gnnforsocialrecommendation}:
\begin{equation}
Loss = \frac{1}{2|T|}\sum_{i,j\in T}(r_{i,j}'-r_{i,j})^2,
\end{equation}
where $|T|$ is the number used in the training dataset, $r_{i,j}'$ is the relationship between entity $v_i$ and $v_j$ predicted by the model and $r_{i,j}$ is the ground truth.

To optimize the objective function, we use Adam \cite{kingma2014adam} as the optimizer in our implementation.
Each time it randomly selects a training instance and updates each model parameter in its negative gradient direction.
In optimizing deep neural network models, overfitting is an eternal problem.
To alleviate this problem, we adopted a dropout strategy \cite{srivastava2014dropout} in the model.
The idea of dropout is to discard some neurons during training randomly.
When updating parameters, only part of them will be updated.
Besides, since the dropout function was disabled during the test, the entire network will be used for prediction.
The whole algorithm framework is shown in algorithm 1.

\section{Experiments} \label{sec: Experiments}

\begin{figure}[t]
\centering
\subfigure[MovieLen]{
\begin{minipage}[t]{1\linewidth}
\centering
\includegraphics[width=1\linewidth]{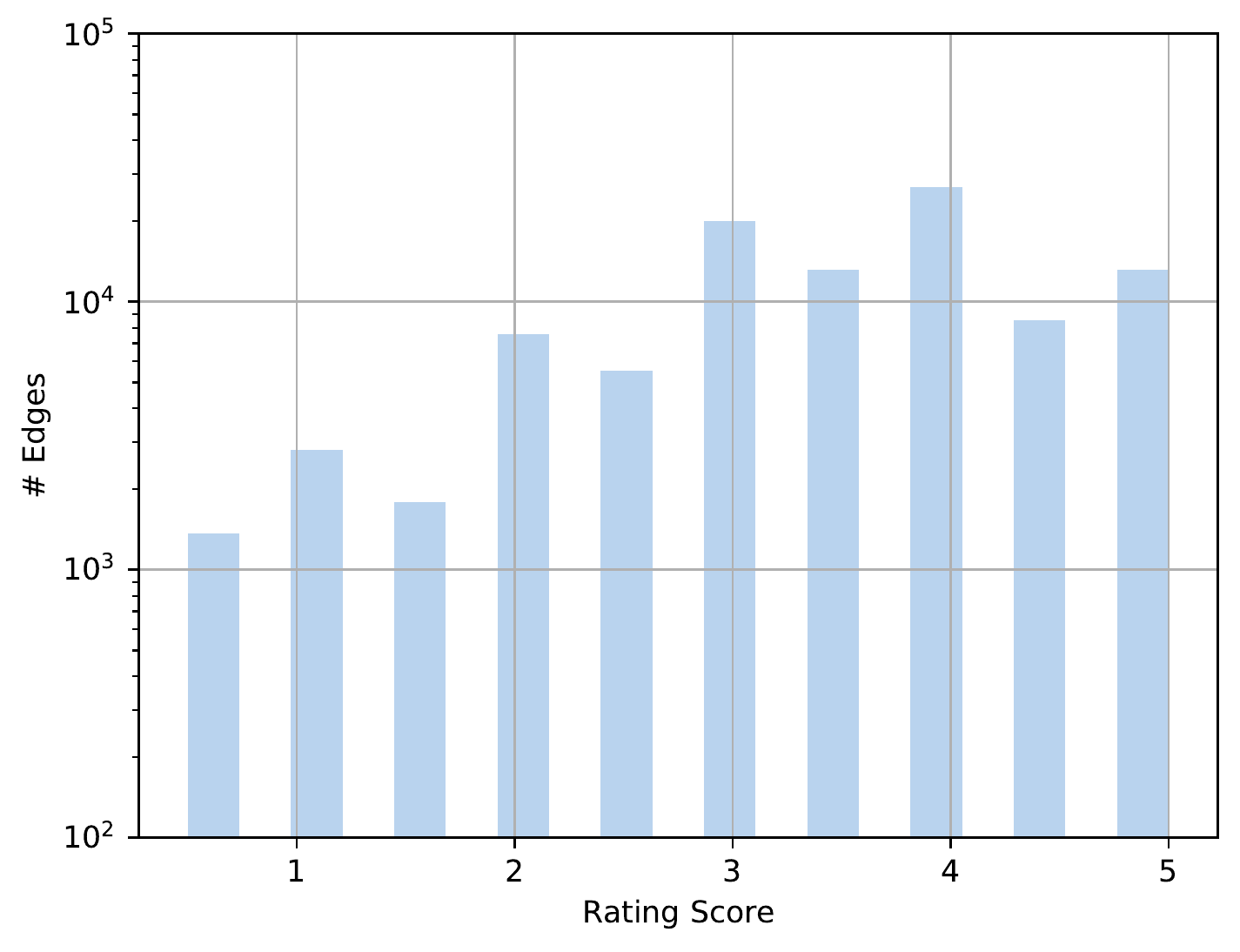}
\end{minipage}
\label{MovieLen-c}
}
\subfigure[AMiner]{
\begin{minipage}[t]{1\linewidth}
\centering
\includegraphics[width=1\linewidth]{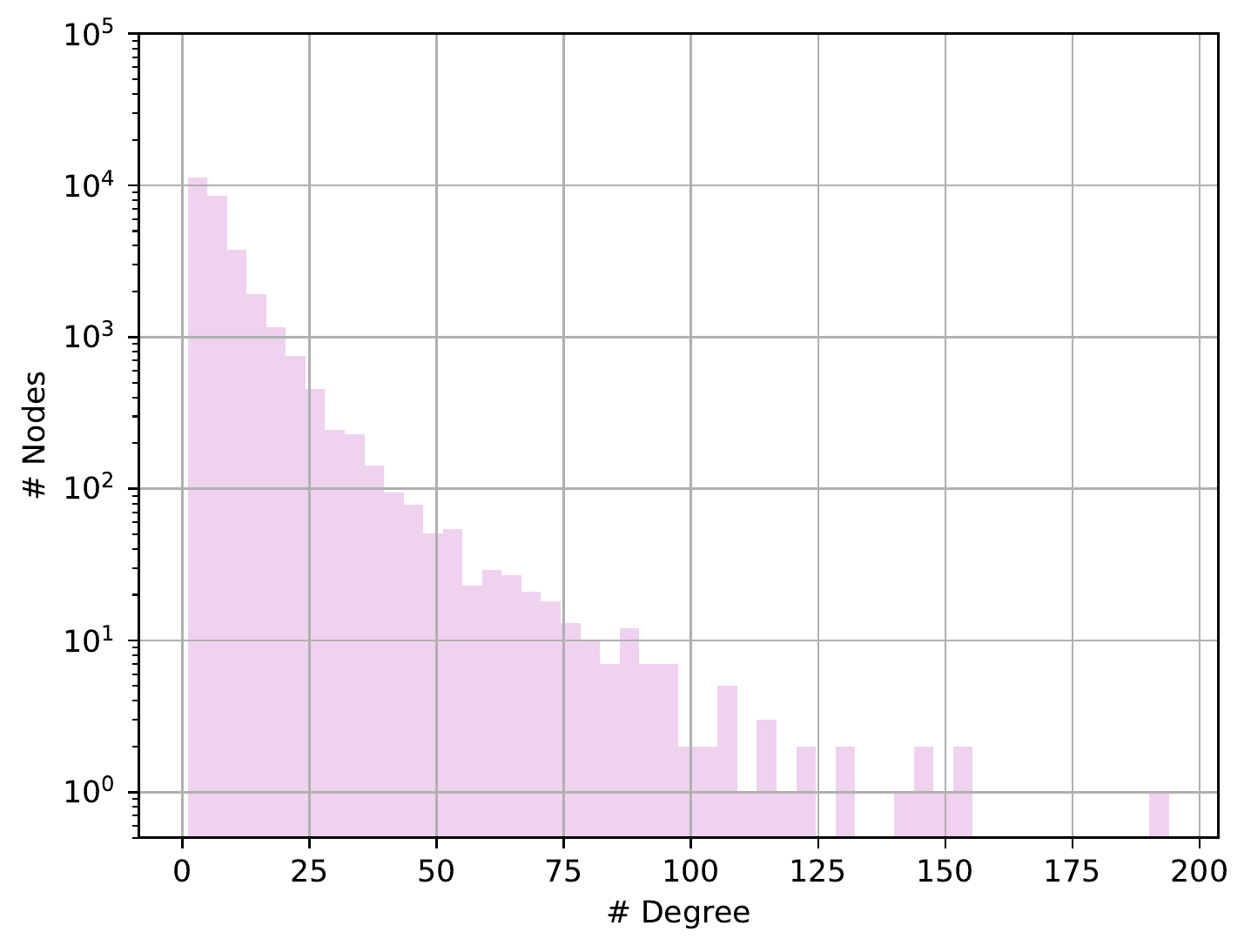}
\end{minipage}
\label{AMiner-c}
}
\caption{Characteristics of the MovieLen and the AMiner dataset.}
\end{figure}

\subsection{Datasets}

In order to verify the effectiveness of HANRec, we performed two tasks: recommendation and link prediction.

For the recommendation task, MovieLens 100k\footnote{https://grouplens.org/datasets/movielens/100k/} was used, a common data set in the field of movie recommendation.
It consists of 10,334 nodes, including 610 users and 9724 movies, and 100,836 edges.
It is worth mentioning that the edges here all indicate the user's rating of the movie, and there is no edge connection between users or between movies.
MovieLens also includes the movie's genre attributes and the timestamp of the rating.
Movie categories include 18 categories such as Action, Drama, Fantasy.
Each movie can have multiple unique category attributes.
For example, Batman (1989) has three category attributes: Action, Crime, and Thriller.
We initialize representation embeddings for these 18 genres.
The initial embedding of each movie is the average of the genre's embeddings in the movie.
For the MovieLen dataset, our model focuses on three kinds of embeddings, including user embedding $h_i^0$, where $i$ belongs to the user index, movie embedding $h_j^0$, which is composed of the embedding of movie genres, and $j$ belongs to the movie index, and opinion embedding $e_{i,j}$.
They are initialized randomly and learn together during the training phase.
Since the original features are extensive and sparse, we do not use one-key vectors to represent each user and item.
By embedding high-dimensional sparse features into the low-dimensional latent space, the model can be easily trained \cite{he2017neural}.
The opinion embedding matrix $e_{i,j}$ depends on the system's rating range.
For example, for the MovieLen dataset, which is a 5-star rating system and 0.5 as an interval, the opinion embedding matrix e contains nine different embedding vectors to represent \{0.5, 1, 1.5, 2, 2.5, 3, 3.5, 4, 4.5, 5\} in the score.
Fig. \ref{MovieLen-c} shows the rating score distribution of edges in the MovieLen dataset.
We can clearly see fewer edges with lower rating scores, and there are more edges with rating scores from 3 to 4.

For the link prediction task, we use the AMiner dataset\footnote{https://www.aminer.cn/aminernetwork} \cite{tang2008arnetminer}.
We construct the relationship between an author and one paper when the author published this paper.
We also generate the relationships between authors when they are co-authors and generate the relationships between papers when there are citation relationships.
The authors' initial attribute contains research interests, published papers, and the total number of citations. \cite{stallings2013determining}.
The title and abstract can represent the initial attribute of papers.
In this author-paper-network, we treat the weight of edges as binary.
In the experiment part, we selected all papers from the famous venue\footnote{1. IEEE Trans. Parallel Distrib. Syst; 2. STOC; 3. IEEE Communications Magazine; 4. ACM Trans. Graph; 5.CHI; 6. ACL; 7. CVPR; 8. WWW} in eight research topics \cite{dong2017metapath2vec} and all the relative authors who published these papers.
Based on this, we derive a heterogeneous attributed network from the academic network of the AMiner dataset.
It consists of 29,059 nodes, including 16,604 authors and 12,455 papers with eight labels, and 124,626 edges representing 62,115 coauthor, 31,251 citation, and 31,263 author-paper relationships. We treat authors' and papers' text descriptions as node attributes and transform them into vectors by Doc2vec in the experiments.
Fig. \ref{AMiner-c} shows the node distribution of the AMiner dataset.
We can find that most nodes have few neighbors, but there are still some super nodes whose number of neighbors exceeds 100.
In this case, it is more important to distinguish the influence of different neighbors effectively.

\subsection{Metrics}

We apply two conventional evaluation metrics to evaluate the performance of different models for recommendation: Mean Absolute Error (MAE) and Root Mean Squared Error (RMSE):

\begin{equation}
MAE=\frac{1}{m}\sum\limits_{i = 1}^{m} \left (p_i-a_i \right )
\end{equation}

\begin{equation}
RMSE=\sqrt{\frac{1}{m}\sum\limits_{i = 1}^{m} (p_i-a_i)^2}
\end{equation}

\begin{center}
   \textit{a = actual target}, \textit{p = predict target}\\ 
\end{center}

Smaller values of MAE and RMSE indicate better predictive accuracy.
For the link prediction task, Accuracy and AUC (Area Under the Curve) are used to quantify the accuracy. Higher values of Accuracy and AUC indicate better predictive accuracy.

\subsection{Baselines}

The methods in our comparative evaluation are as follows:
\begin{itemize}
\item \textbf{Doc2Vec} \cite{le2014distributed} is the Paragraph Vectors algorithm that embeds the text describing objects in a distributed vector using neural network models. Here we use Doc2Vec to process the text describing authors and papers' research interests in the link prediction task to obtain the initialization embedding of authors and papers.
\item \textbf{DeepWalk} \cite{perozzi2014deepwalk} uses random walk to sample nodes in the graph to get the node embedding. As for the relevant parameters, we refer to the original paper. We set num-walks as 80, walk-length as 40, and window-size as 10.
\item \textbf{LINE} \cite{tang2015line} minimizes a loss function to learn embedding while preserving the first and the second-order neighbors' proximity among nodes. We use LINE (1st+2nd) as overall embeddings. The number of negative samples is set as 5, just the same as the original paper.
\item \textbf{Node2Vec} \cite{grover2016node2vec} adopts a biased random walk strategy and applies Skip-Gram to learn node embedding. We set $p$ is 0.25 and $q$ as 4 here.
\item \textbf{SoRec} \cite{ma2008sorec} performs co-factorization on the user-item rating matrix and user-user social relations matrix. We set the parameters as the same as the original paper. $\lambda_C$ = 10, and $\lambda_U$ = $\lambda_V$ = $\lambda_Z$ = 0.001.
\item \textbf{GATNE} \cite{cen2019representation} provides an embedding method for large heterogeneous network. In the two datasets used here, the edge type is 1, so the edge embedding type in GATNE is set to 1.
\item \textbf{GraphRec} \cite{fan2019graph} jointly captures interactions and opinions in the user-item graph. For the MovieLen dataset here, there are no user-to-user and movie-to-movie interactions, so only item aggregation and user aggregation of the original paper are used.
\item \textbf{HANRec} is our proposed framework, which makes full use of the heterogeneity and attribute of the network and uses the attention mechanism to provide better recommendations.
\end{itemize}

\begin{table*}[t]
\renewcommand\arraystretch{1.2}
\centering
\caption{Results comparison of the recommendation. MAE/RMSE are evaluation metrics.}
\scalebox{0.8}{
\begin{tabular}{c|c|c|c|c|c|c|c}
\hline
Train   Edges & DeepWalk      & LINE          & Node2Vec      & SoRec         & GATNE         & GraphRec      & HANRec                 \\ \hline
30\%          & 0.8514/1.0578 & 0.8614/1.0687 & 0.8519/1.0317 & 0.8278/1.0324 & 0.7923/0.9848 & 0.7824/0.9748 & \textbf{0.7751/0.9624} \\ \hline
40\%          & 0.8258/1.0014 & 0.8413/1.0342 & 0.8202/0.9832 & 0.7891/0.9936 & 0.7774/0.9607 & 0.7528/0.9555 & \textbf{0.7419/0.9355} \\ \hline
50\%          & 0.7869/0.9817 & 0.8017/0.9959 & 0.7928/0.9707 & 0.7417/0.9615 & 0.7319/0.9487 & 0.7347/0.9379 & \textbf{0.7128/0.9132} \\ \hline
60\%          & 0.7521/0.9678 & 0.7758/0.9816 & 0.7498/0.9601 & 0.7314/0.9504 & 0.7217/0.9215 & 0.7191/0.9189 & \textbf{0.6981/0.9047} \\ \hline
70\%          & 0.7331/0.9497 & 0.7525/0.9607 & 0.7345/0.9427 & 0.7257/0.9407 & 0.7047/0.9158 & 0.7055/0.9107 & \textbf{0.6823/0.8925} \\ \hline
80\%          & 0.7250/0.9418 & 0.7332/0.9511 & 0.7273/0.9395 & 0.7192/0.9339 & 0.6954/0.9057 & 0.6966/0.9071 & \textbf{0.6753/0.8855} \\ \hline
90\%          & 0.7148/0.9375 & 0.7241/0.9403 & 0.7068/0.9189 & 0.6957/0.9048 & 0.6824/0.8907 & 0.6807/0.8827 & \textbf{0.6673/0.8681} \\ \hline
\end{tabular}
}
\label{recommendation}
\end{table*}

\begin{table*}[]
\renewcommand\arraystretch{1.2}
\centering
\caption{Results comparison of the link prediction. AUC/Accuracy are evaluation metrics.}
\scalebox{0.8}{
\begin{tabular}{c|c|c|c|c|c|c|c}
\hline
Train   Edges & Doc2Vec       & DeepWalk      & LINE          & Node2Vec      & SoRec         & GraphRec      & HANRec                 \\ \hline
30\%          & 0.8074/0.7335 & 0.8693/0.8206 & 0.7911/0.7214 & 0.8808/0.8341 & 0.9228/0.8547 & 0.9241/0.8613 & \textbf{0.9612/0.8982} \\ \hline
40\%          & 0.8158/0.7407 & 0.8873/0.8345 & 0.8017/0.7275 & 0.8879/0.8410 & 0.9317/0.8602 & 0.9333/0.8752 & \textbf{0.9647/0.9032} \\ \hline
50\%          & 0.8243/0.7505 & 0.8948/0.8415 & 0.8155/0.7357 & 0.9023/0.8531 & 0.9394/0.8739 & 0.9517/0.8907 & \textbf{0.9679/0.9101} \\ \hline
60\%          & 0.8395/0.7612 & 0.9105/0.8631 & 0.8209/0.7398 & 0.9201/0.8617 & 0.9534/0.8907 & 0.9571/0.8988 & \textbf{0.9758/0.9223} \\ \hline
70\%          & 0.8548/0.7765 & 0.9358/0.8758 & 0.8288/0.7451 & 0.9393/0.8826 & 0.9627/0.9015 & 0.9673/0.9056 & \textbf{0.9807/0.9409} \\ \hline
80\%          & 0.8702/0.7893 & 0.9501/0.8921 & 0.8371/0.7473 & 0.9513/0.8901 & 0.9708/0.9077 & 0.9728/0.9099 & \textbf{0.9881/0.9514} \\ \hline
90\%          & 0.8771/0.7973 & 0.9573/0.8967 & 0.8429/0.7506 & 0.9584/0.8972 & 0.9782/0.9103 & 0.9792/0.9156 & \textbf{0.9912/0.9611} \\ \hline
\end{tabular}
}
\label{link}
\vspace{-5mm}
\end{table*}

\subsection{Training Details}

For the task of recommendation, we use the MovieLen dataset.
The recommendation task's goal is to predict the rating score of one user to one movie.
$x$\% of the scoring records are used for training, and the remaining (100-$x$)\% are used for testing.
In this task, HANRec will output a value between 0.5 and 5.0, indicating the user's possible rating for the movie.
The closer the score is to the ground truch, the better the performance of the model.
For the task of link prediction, we use the AMiner dataset.
The link prediction task's goal is to predict whether there is an edge between two given nodes.
First, we randomly hide (100-$x$)\% of the edges in the original graph to form positive samples in the test set.
The test set also has an equal number of randomly selected disconnected links that servers as negative samples.
We then use the remaining $x$\% connected links and randomly selected disconnected ones to form the training set.
HANRec outputs a value from 0 to 1, indicating the probability that there are edges between entities.
For these two tasks, the value of $x$ ranges from 30 to 90, and the interval is 10.
It is worth noting that when $x$ is low, we call it a cold start problem \cite{schafer2007collaborative}, which is discussed in detail in the section \ref{cold-start}.

The proposed HANRec is implemented on the basis of Pytorch 1.4.0\footnote{https://pytorch.org/}.
All multilayer perceptrons have a three-layer linear network and prelu activation function by default.
The embedding size $d$ used in the model and the batch size are all set to 128.
The learning rate are searched in [0.001, 0.0001, 0.00001], and the Adam algorithm \cite{kingma2014adam} is used to optimize the parameters of HANRec.
The performance results were recorded on the test set after 2000 iterations on the MovieLen dataset and 20000 iterations on the AMiner dataset.
The parameters for the baseline algorithms were initialized as in the corresponding papers and were then carefully tuned to achieve optimal performance.

\subsection{Main Results}

We can first see that no matter which task it is, as the proportion of the training data set increases, the effect of HANRec on the test set gets better and better.
Then we focus on the recommendation performance of all methods.
Table \ref{recommendation} shows the overall rating prediction results (MAE and RMSE) among different recommendation methods on MovieLen datasets.
We find that no matter how much $x$ is, the performance of DeepWalk, LINE, and Node2Vec, which are based on the walking in the graph, is relatively low.
As a matrix factorization method that leverages both the rating and social network information, SoRec is slightly better than the previous three.
GATNE's embedding of edge heterogeneity and GraphRec's social aggregation cannot be fully utilized in this case.
Nevertheless, these two methods' performance is generally better than the previous methods, which proves the effectiveness of deep neural networks on the recommendation task.
The model we proposed, HANRec, can fully connect users, thus providing more guidance for the recommendation, thus showing the best performance on the recommendation task.

Then we evaluate HANRec's performance on the link prediction accuracy and compare HANRec with other link prediction algorithms mentioned above.
Table \ref{link} demonstrates that the performance of Doc2Vec, which does not use graph information and only uses the initial information of each node, is the worst.
This shows that the graph structure is essential in the link prediction task, and the initial features of the nodes are challenging to provide enough reference for the link prediction task.
Methods that use graph structure information, such as DeepWalk, LINE, Node2Vec, and SoRec, have higher performance than Doc2Vec, indicating that in this case, the graph structure information is more important than the initial information of nodes.
The effect of GATNE, which uses graph heterogeneity, is better than the previously mentioned methods, which shows that graph heterogeneity can also provide some reference for the link prediction task.
The performance of our method HANRec in the link prediction task far exceeds other methods, which further shows that the framework can fully consider the homogeneity and heterogeneity of the graph and use the attention mechanism to generate high-quality embeddings for nodes.
Further investigations are also conducted to better understand each component's contributions in HANRec in the following subsection.

In conclusion, we can know from these results:
(1) graph neural networks can boost the performance in the recommendation and the link prediction tasks;
(2) using the heterogeneity of graphs can make full use of the information of the network, thereby improving the performance on these two tasks;
(3) our proposed HANRec achieves the state-of-art performance in the MovieLen and the AMiner datasets for both tasks.

\begin{figure}[t]
\centering
\subfigure[Cold-Start: MovieLen-MAE]{
\begin{minipage}[t]{0.47\linewidth}
\centering
\includegraphics[width=1\linewidth]{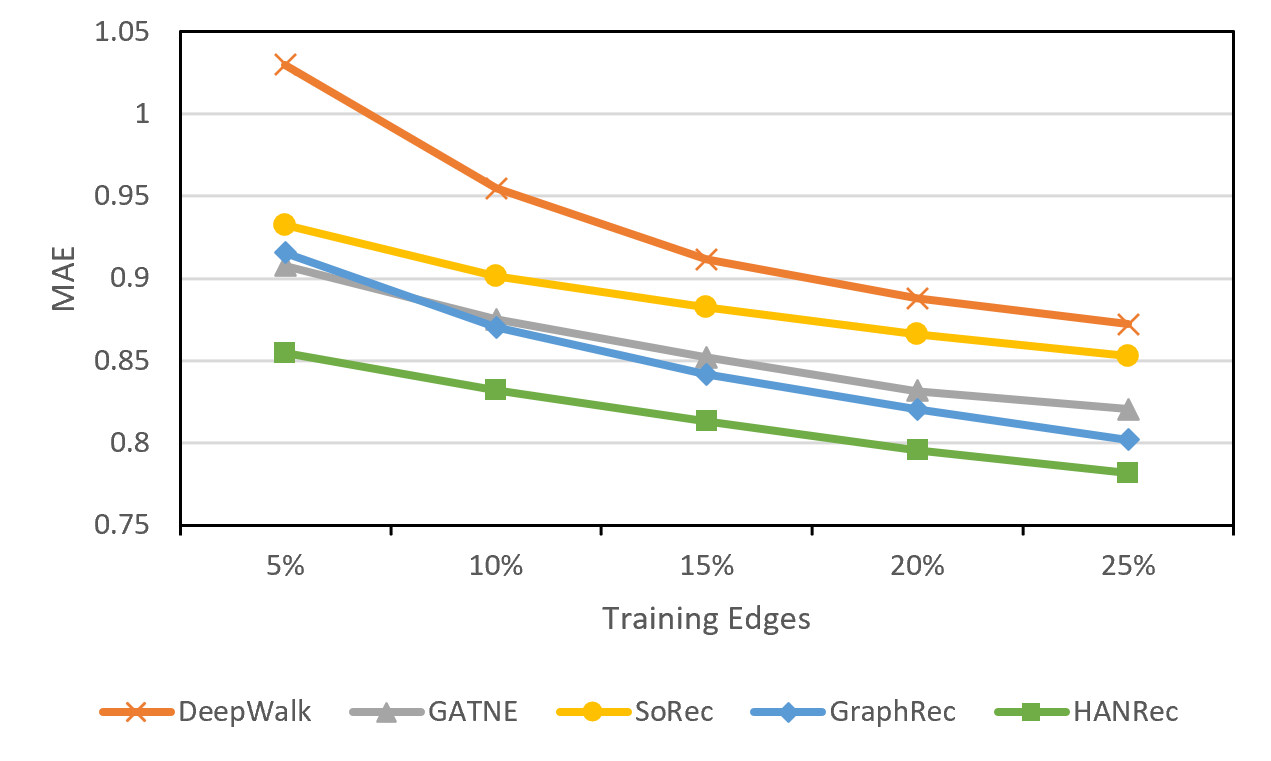}
\label{cold-MoiveLen}
\end{minipage}
}
\subfigure[Cold-Start: AMiner-AUC]{
\begin{minipage}[t]{0.47\linewidth}
\centering
\includegraphics[width=1\linewidth]{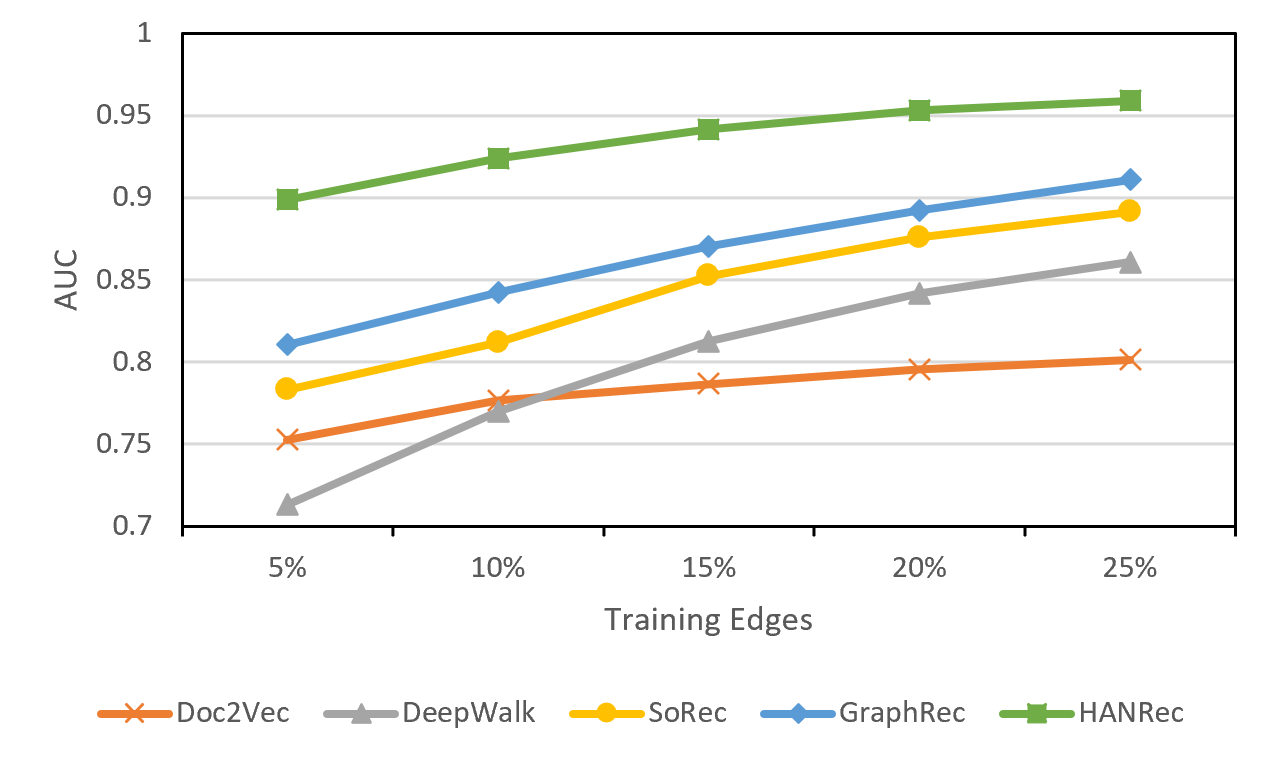}
\label{cold-AMiner}
\end{minipage}
}
\caption{Cold Start Problem analysis on the MovieLen and AMiner datasets.}
\end{figure}

\subsection{Cold Start Problem}
\label{cold-start}
The cold start problem \cite{schafer2007collaborative} refers to the difficulty of the recommendation system to give high quality recommendations when there is insufficient data.
According to \cite{bobadilla2012collaborative}, the cold start problem can be divided into three categories:
(1) User cold start: how to make personalized recommendations for new users;
(2) Item cold start: how to recommend a new item to users who may be interested in it;
(3) System cold start: how to design a personalized recommendation system on a newly developed website (no users, user behavior, only partial item information), so that users can experience personalized recommendations when the website is released.

In this paper, the cold start problem belongs to the third category, which can be defined as how to make high-quality recommendations based on the information of a small number of edges (rating scores or relationships).
We continuously adjusted the proportion of available information in the two datasets, that is, the proportion of training set x, from 5\% to 25\%, and obtained the performance results of a series of methods.
When the ratio of the training edges is from 5\% to 25\%, Fig. \ref{cold-MoiveLen} and \ref{cold-AMiner} show the performance of different models in the MoveLen and the AMiner datasets.
Since the methods based on the random walk, such as Node2Vec and LINE, have a similar performance with DeepWalk, we only report the results of DeepWalk here.

Both Fig. \ref{cold-MoiveLen} and Fig. \ref{cold-AMiner} reflect that HANRec performs the best in various cases.
Specifically, when the proportion of training edges is 5\%, HANRec outperforms other methods the most.
This reflects the superiority of the connect method we designed: when there are very few known edges, the entity's neighbor information is incomplete.
The embedding obtained by other methods by aggregating neighbor information does not fully integrate this neighbor information.
By connecting potential neighbors, HANRec overcomes the disadvantage of few known edges to a certain extent and is suitable for alleviating the difficulty of generating high-quality recommendations during cold start.
It is worth noticing that as shown in Fig. \ref{cold-AMiner}, Doc2Vec, the recommended method based on entity features, is better than DeepWalk as the first, which is based on the random walk.
As the proportion of the training edges increases, DeepWalk gradually overtakes Doc2Vec.
This is intuitive: because DocVec does not use the graph's structural information, it only makes recommendations based on each entity's feature.
The increase in the number of training edges has relatively little effect on the performance improvement of Doc2Vec.
While DeepWalk can capture more neighbor information by random walking, improve the generated embedding quality, and make better recommendations.

\begin{figure}[t]
\centering
\subfigure[Ablation Study: MovieLen-MAE]{
\begin{minipage}[t]{0.47\linewidth}
\centering
\includegraphics[width=1\linewidth]{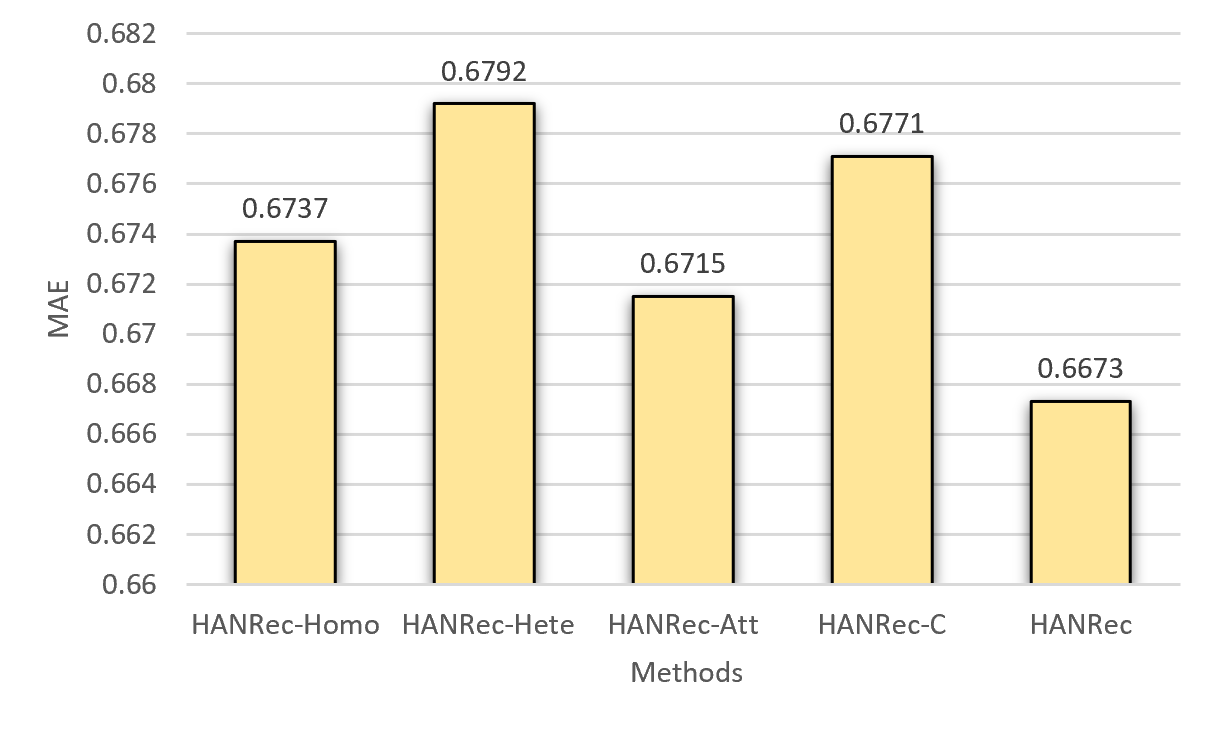}
\end{minipage}
}
\subfigure[Ablation Study: MovieLen-RMSE]{
\begin{minipage}[t]{0.47\linewidth}
\centering
\includegraphics[width=1\linewidth]{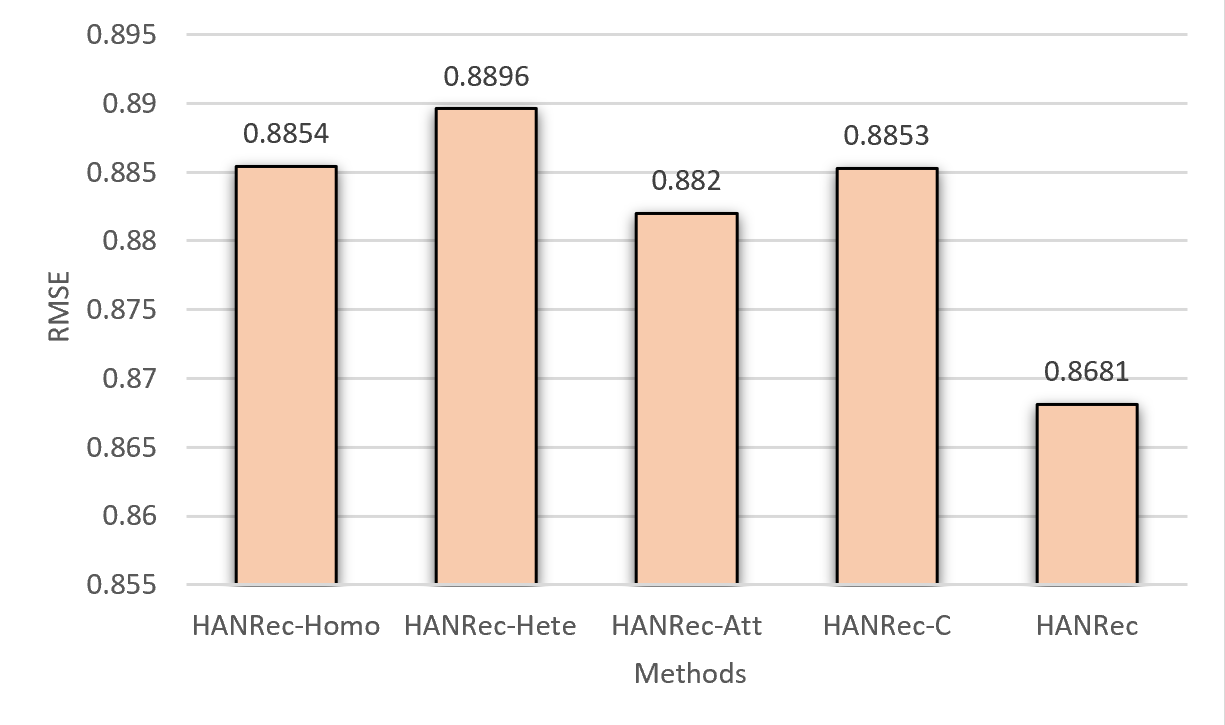}
\end{minipage}
}
\\
\subfigure[Ablation Study: AMiner-AUC]{
\begin{minipage}[t]{0.47\linewidth}
\centering
\includegraphics[width=1\linewidth]{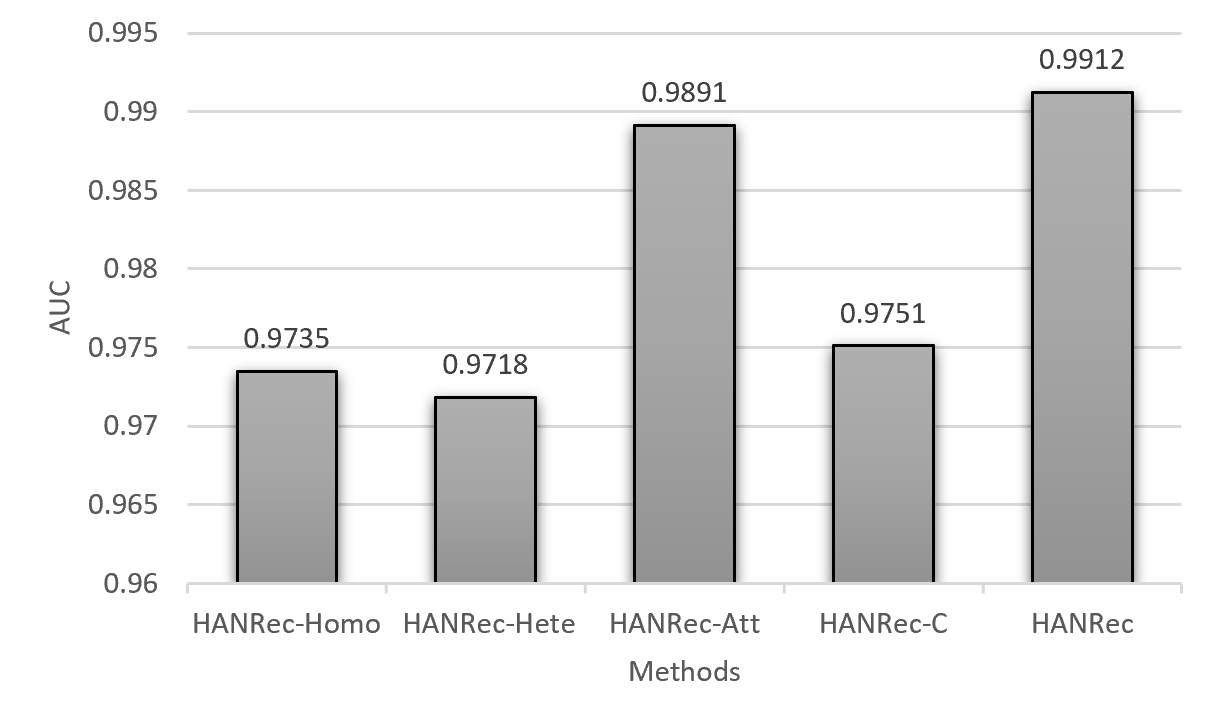}
\end{minipage}
}
\subfigure[Ablation Study: AMiner-Accuracy]{
\begin{minipage}[t]{0.47\linewidth}
\centering
\includegraphics[width=1\linewidth]{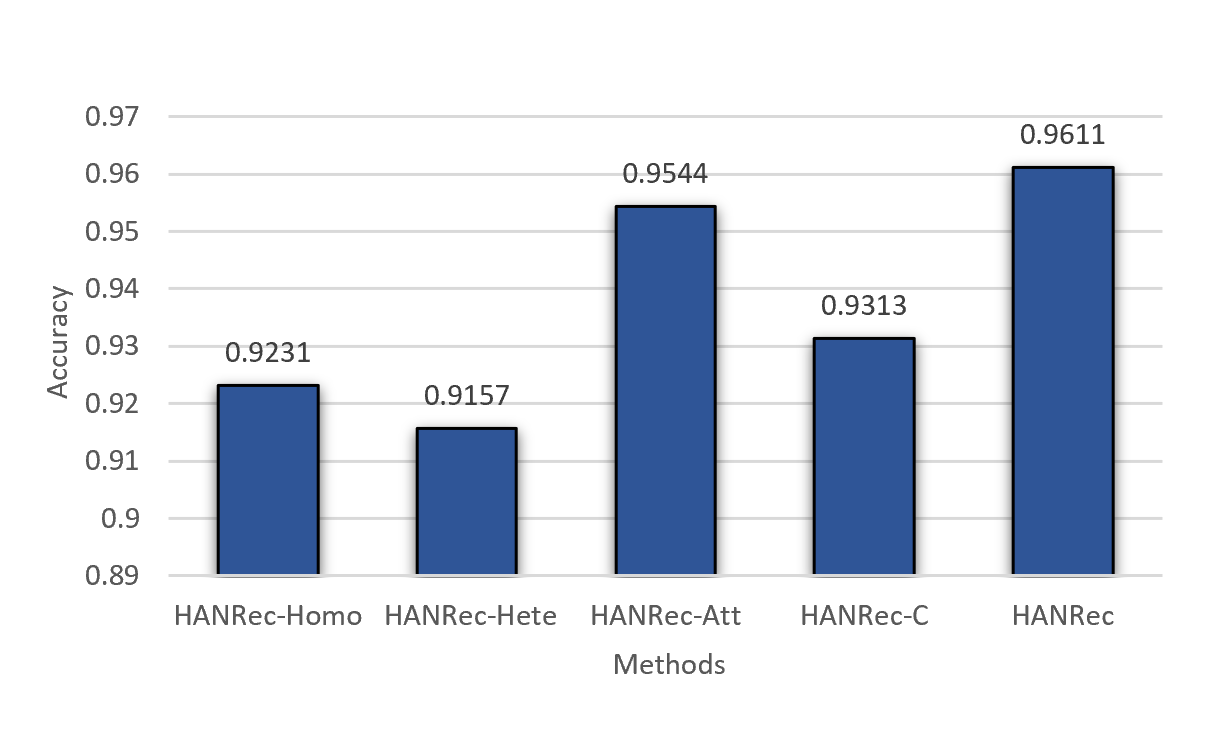}
\end{minipage}
}
\caption{Effect of each component on the MovieLen and the AMiner dataset.}
\label{ablation_component}
\vspace{-5mm}
\end{figure}

\begin{figure}[t]
\centering
\subfigure[Connecting Strategies: MovieLen-MAE]{
\begin{minipage}[t]{0.47\linewidth}
\centering
\includegraphics[width=1\linewidth]{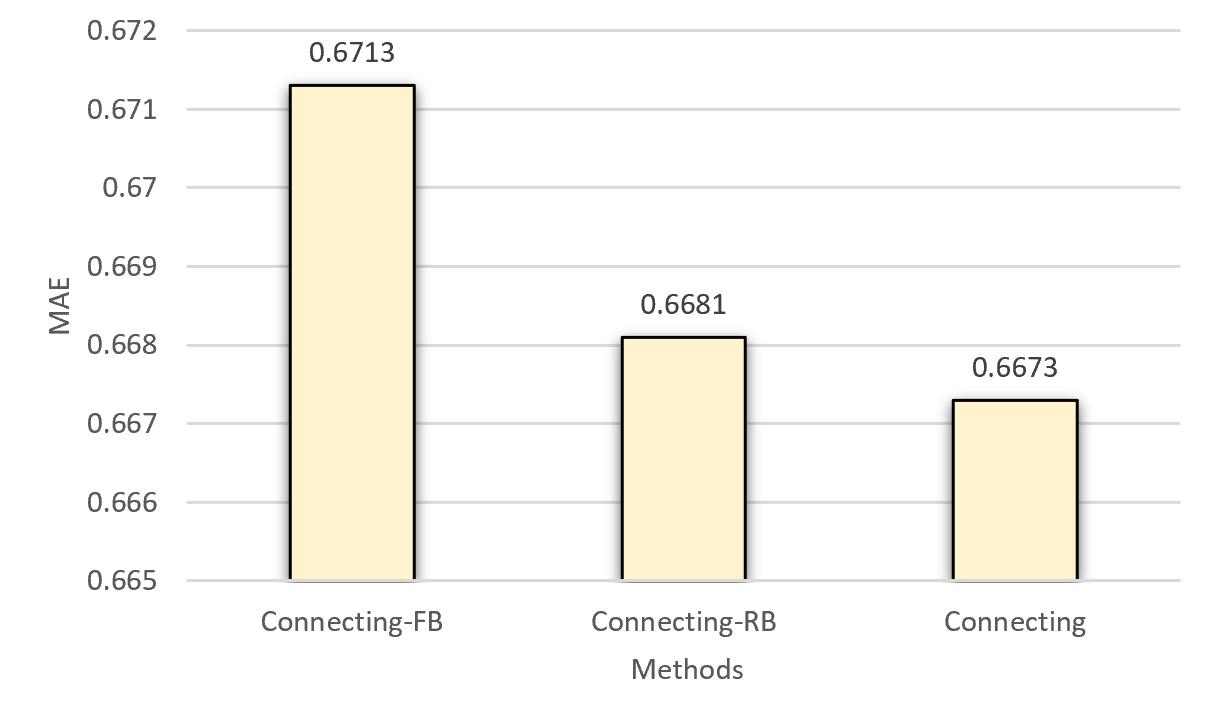}
\end{minipage}
}
\subfigure[Connecting Strategies: MovieLen-RMSE]{
\begin{minipage}[t]{0.47\linewidth}
\centering
\includegraphics[width=1\linewidth]{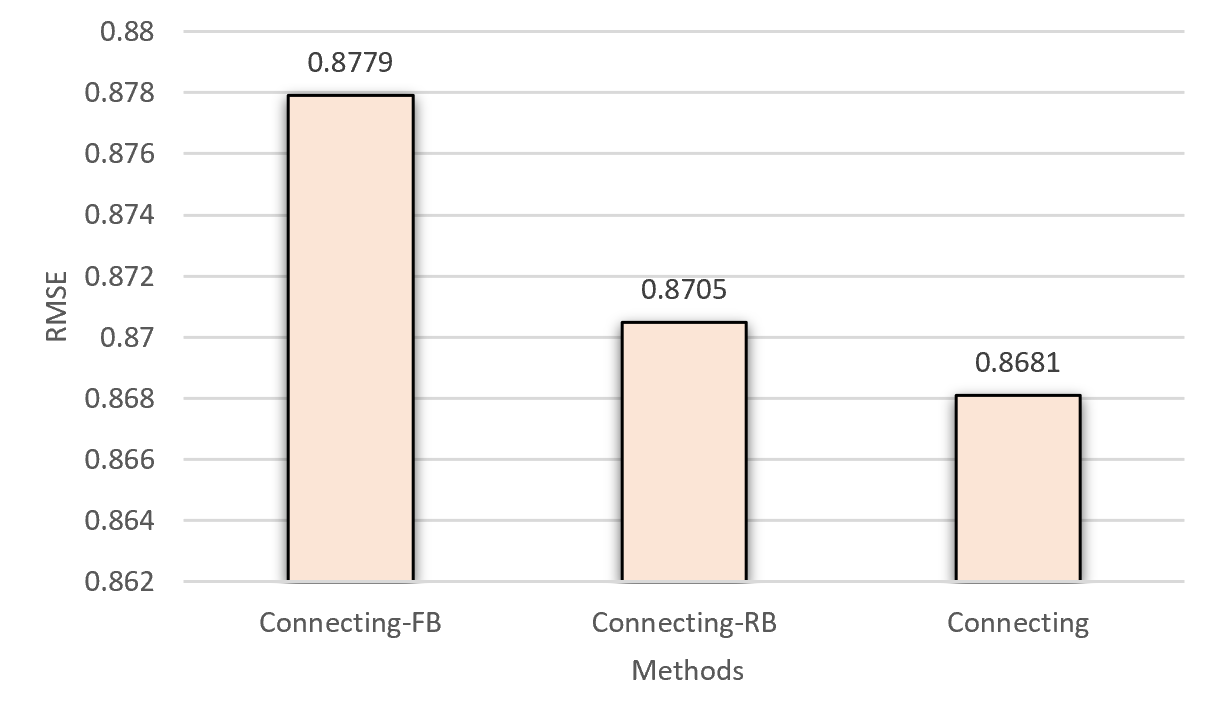}
\end{minipage}
}
\\
\subfigure[Connecting Strategies: AMiner-AUC]{
\begin{minipage}[t]{0.47\linewidth}
\centering
\includegraphics[width=1\linewidth]{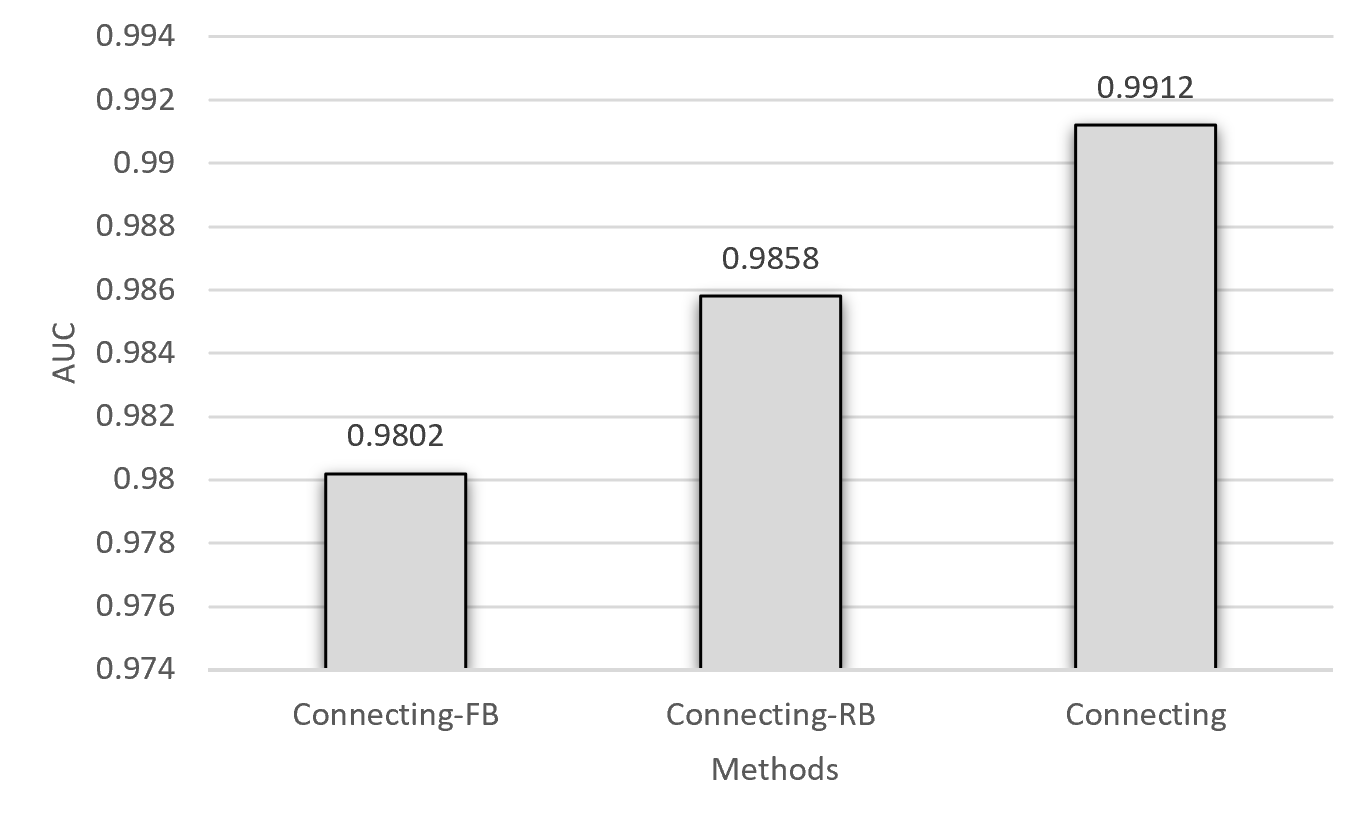}
\end{minipage}
}
\subfigure[Connecting Strategies: AMiner-Accuracy]{
\begin{minipage}[t]{0.47\linewidth}
\centering
\includegraphics[width=1\linewidth]{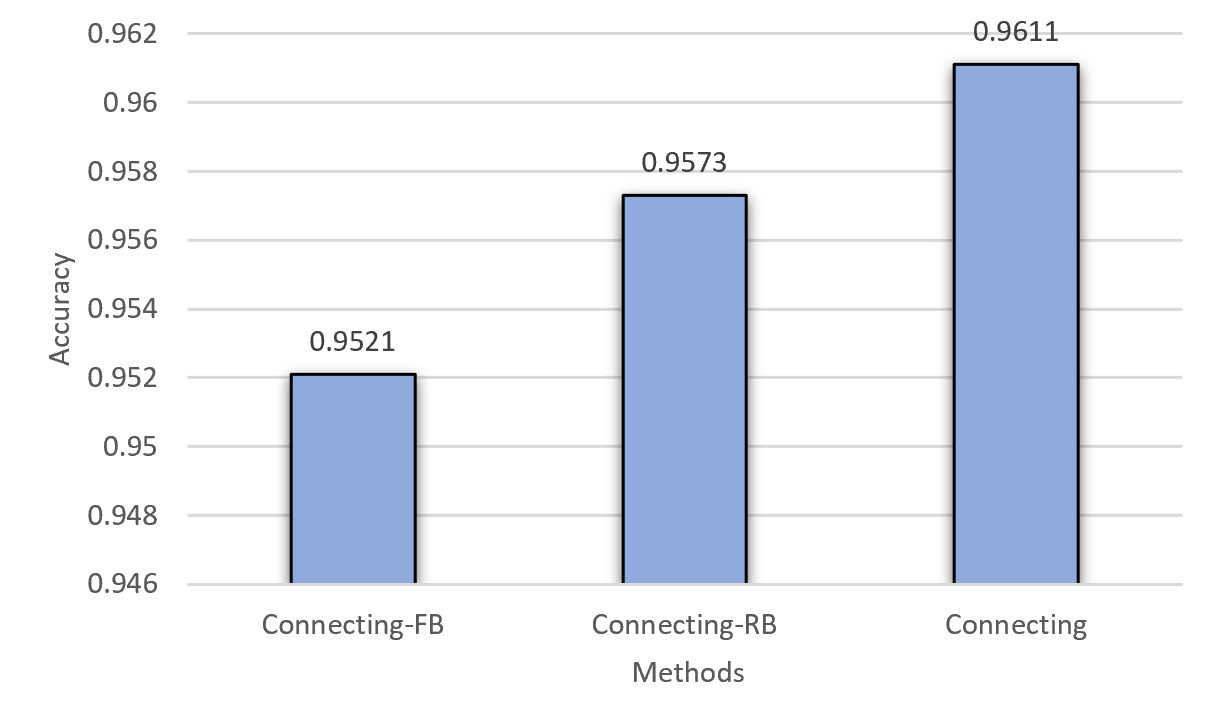}
\end{minipage}
}
\caption{The performance of different connecting methods on the MovieLen and the AMiner dataset.}
\label{ablation_connect}
\vspace{-5mm}
\end{figure}

\subsection{Ablation Study}

We evaluated how each of the components of HANRec affects the results.
We report the results on two datasets after removing a specific part.
HANRec-C, HANRec-Homo, HANRec-Hete, HANRec-Att respectively represent our model without connecting potential neighbors, homogeneous aggregation, heterogeneous aggregation, and the attention mechanism.
HANRec is the complete model we proposed.
It can be seen from Fig. \ref{ablation_component} that regardless of removing the connecting component, homogeneous aggregation, heterogeneous aggregation, or the attention mechanism, the performance of HANRec shows attenuation, and removing the attention mechanism has less impact on performance than the former two.
This makes intuitive sense.
The other three cases will lose much information in the graph, which is not conducive to the entity's high-quality embedding.
It is worth noting that after removing the connecting component, the performance of the model has dropped a lot, indicating that the connection method we designed can provide an essential reference for recommendations.
This also further proves that the connecting component, aggregation methods, and attention mechanism are practical.
With their joint contribution, HANRec can perform well in the recommendation task.

We also explored the performance of different connection strategies.
Connecting-FB represents the feature-based connecting method.
When we explore neighbors with potential relationships, Connecting-FB only uses the feature information of entities on the path, not the relationship information between entities.
In this case, formula \ref{eq:connect} can be rewritten as: $f_{i,j} = MLP(h_k^0 \oplus h_j^0)$.
Connecting-RB means the relation-based connecting method.
Connecting-RB only uses the relationship information between entities on the path, not the feature information of entities.
So formula \ref{eq:connect} can be rewritten as: $f_{i,j} = MLP(e_{i,k} \oplus e_{k,j})$.
Connecting is the complete connecting method we proposed.
As shown in Fig. \ref{ablation_connect}, whether it is a feature-based connection or a relationship-based connection, the model's performance is not as good as the connection method we use.
This is intuitive: for example, in a movie recommendation network, if two users have watched a movie, then both the rating scores of the movie and the characteristics of the movie and users can provide a vital reference for the relationship between users.

\begin{figure}[t]
\centering
\subfigure[Parameter Analysis: MovieLen-MAE]{
\begin{minipage}[t]{0.47\linewidth}
\centering
\includegraphics[width=1\linewidth]{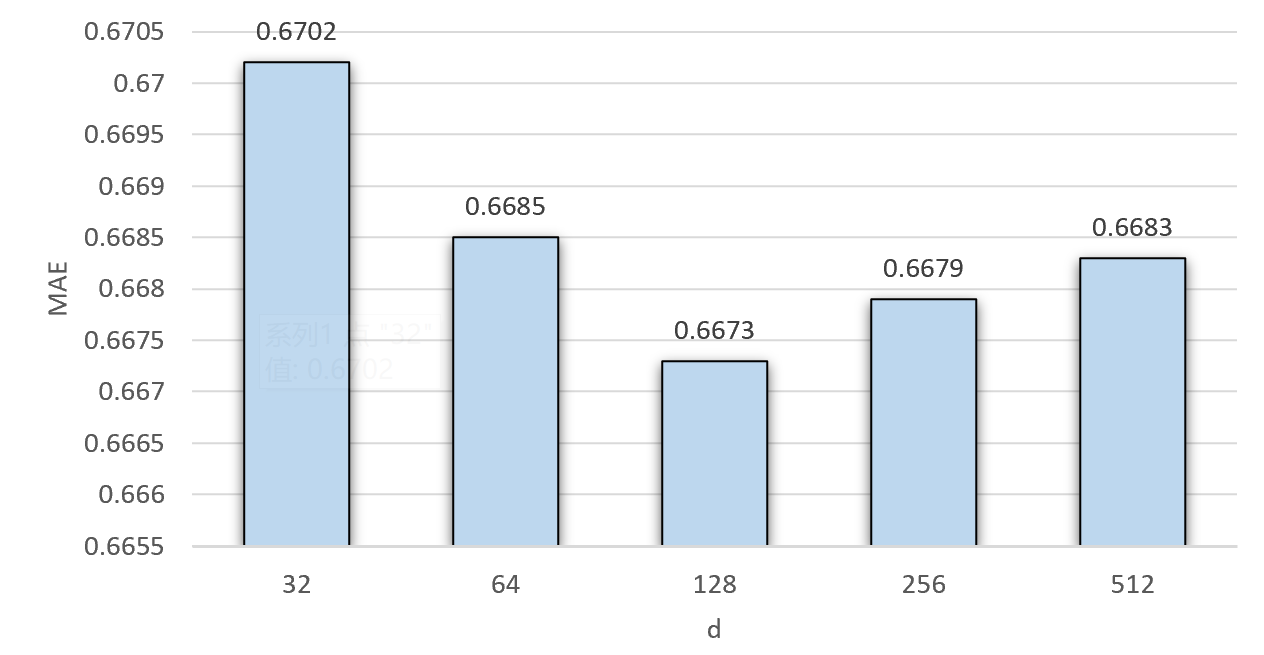}
\end{minipage}
}
\subfigure[Parameter Analysis: MovieLen-RMSE]{
\begin{minipage}[t]{0.47\linewidth}
\centering
\includegraphics[width=1\linewidth]{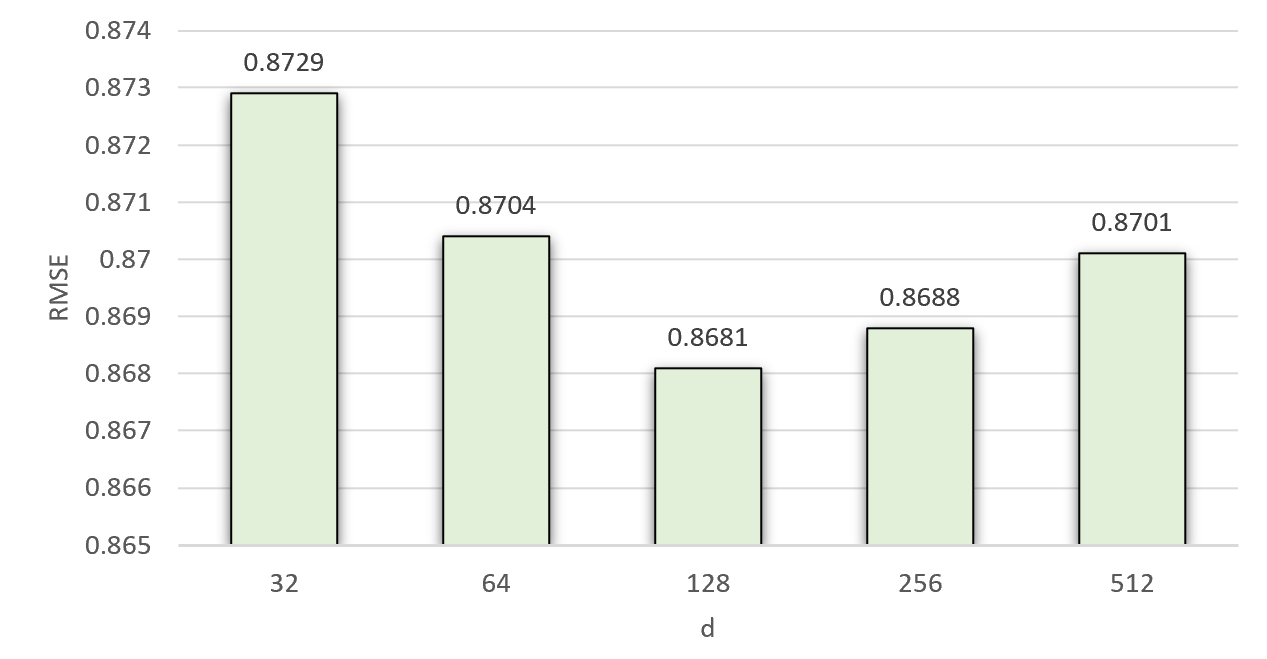}
\end{minipage}
}
\\
\subfigure[Parameter Analysis: AMiner-AUC]{
\begin{minipage}[t]{0.47\linewidth}
\centering
\includegraphics[width=1\linewidth]{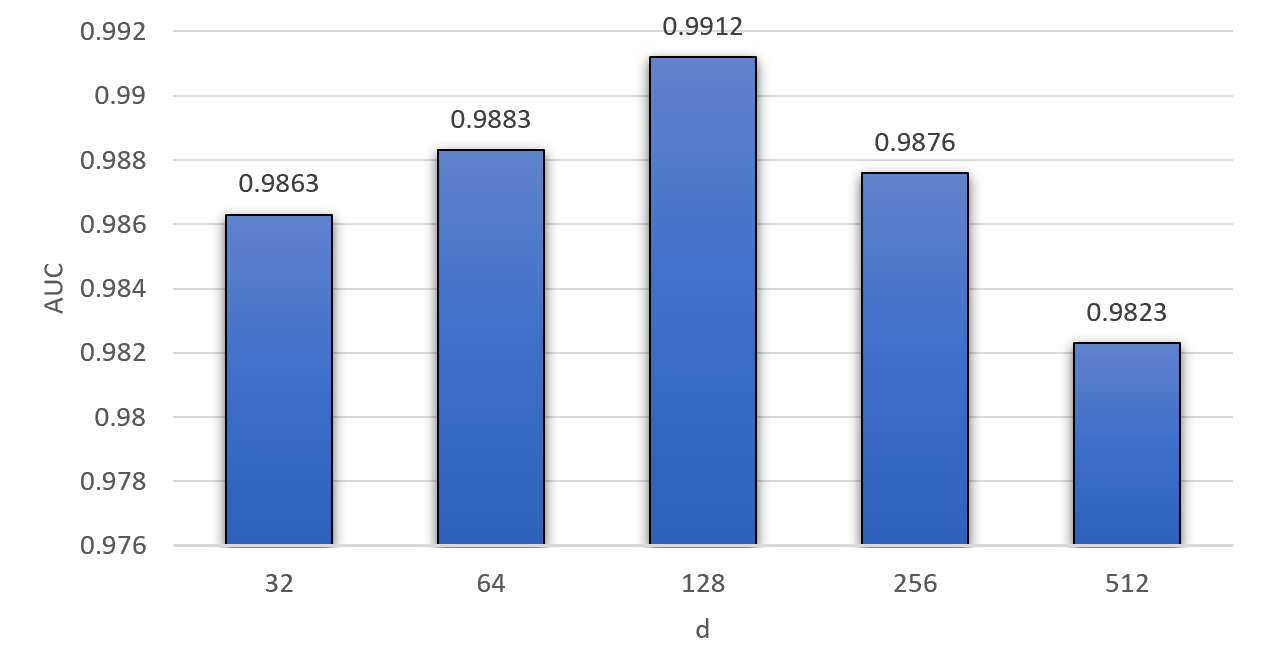}
\end{minipage}
}
\subfigure[Parameter Analysis: AMiner-Accuracy]{
\begin{minipage}[t]{0.47\linewidth}
\centering
\includegraphics[width=1\linewidth]{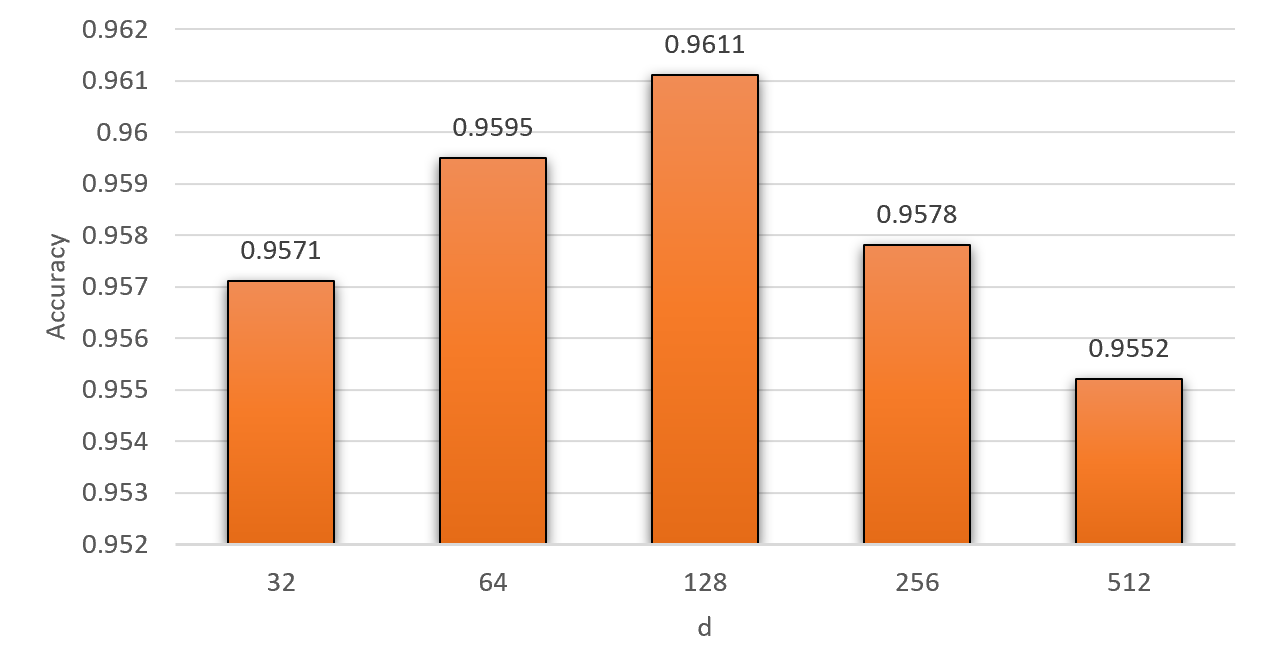}
\end{minipage}
}
\caption{Effect of dimension $d$ on the MovieLen and the AMiner dataset.}
\label{sensitivity}
\end{figure}

\begin{figure}[t]
\centering
\subfigure[MovieLen-Loss]{
\begin{minipage}[t]{0.47\linewidth}
\centering
\includegraphics[width=1\linewidth]{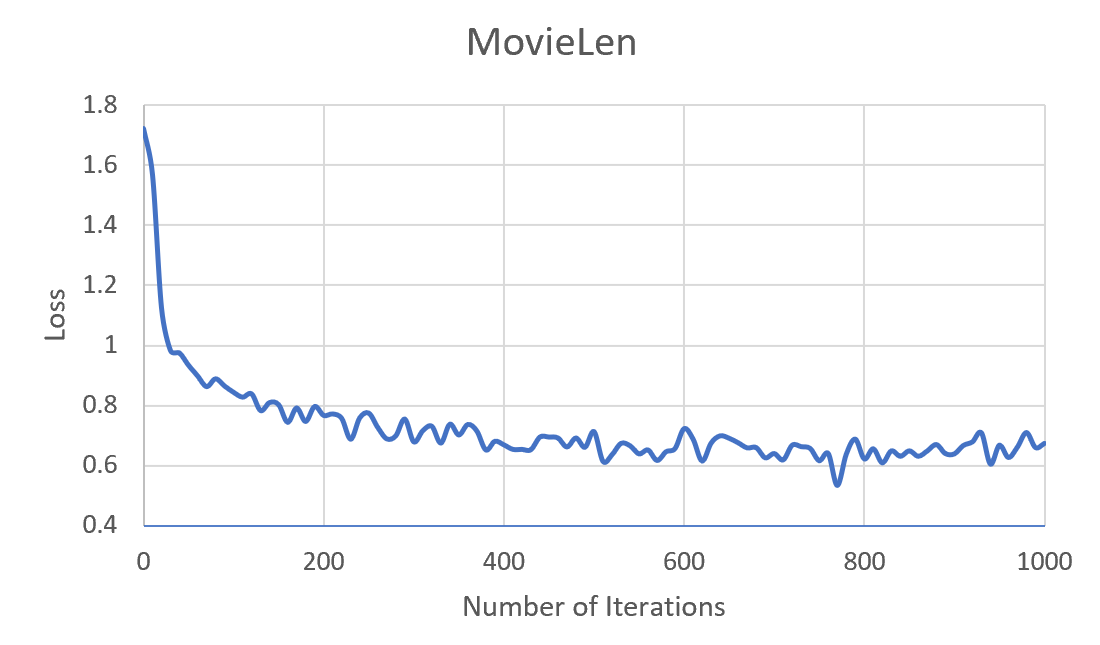}
\end{minipage}
}
\subfigure[AMiner-Loss]{
\begin{minipage}[t]{0.47\linewidth}
\centering
\includegraphics[width=1\linewidth]{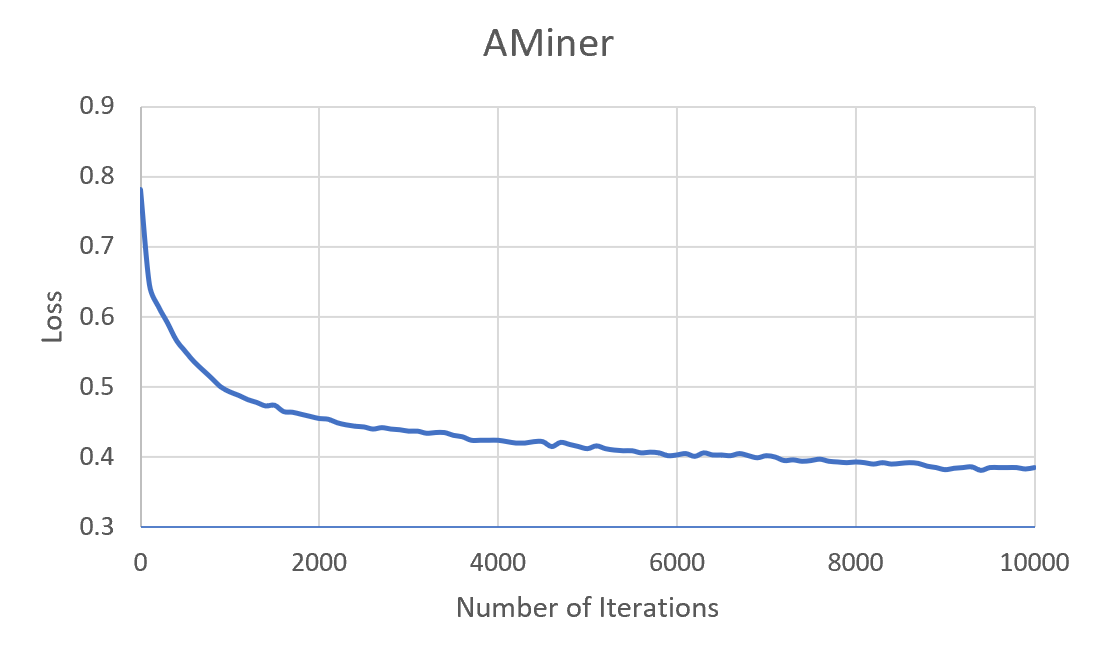}
\end{minipage}
}
\caption{Performance with respect to the number of iterations on the MovieLen and the AMiner dataset.}
\label{loss}
\end{figure}

\subsection{Parameter Analysis}

In this part, we analyze the effect of embedding dimension $d$ of the entity's latent representation $p_i$ on the performance of HANRec. 
Fig. \ref{sensitivity} presents the performance comparison of the embedding dimension on the MovieLen and the AMiner dataset. 
In general, with the increase of the embedding dimension, the performance first increases and then decreases. 
When expanding the embedding dimension from 32 to 128 can improve the performance significantly. 
However, with the embedding dimension of 256, HANRec degrades the performance.
It demonstrates that using a large number of the embedding dimension has powerful representation. Nevertheless, if the embedding dimension is too large, the complexity of our model will significantly increase. 
Therefore, we need to find a proper length of embedding to balance the trade-off between the performance and the complexity. 

We also study the performance change w.r.t. the number of iterations when the learning rate is 0.001, and the training ratio is 90\%.
As shown in Fig. \ref{loss}, we can see that the proposed model has a fast convergence rate, and about 400 iterations are required for the MovieLen dataset, while about 8000 iterations are required for the AMiner dataset.

\section{Conclusion} 
\label{sec: Conclusion}
In this paper, we propose a heterogeneous attributed network framework with a connecting method, called HANRec, to address the recommendation task in heterogeneous graph.
By gathering multiple types of neighbor information and using the attention mechanism, HANRec efficiently generates embeddings of each entity for downstream tasks.
The experiment results on two real-world dataset can prove HANRec outperform state-of-the art models for the task of recommendation and link prediction.

\bibliography{ref}

\end{document}